\pdfoutput=1

\RequirePackage{fix-cm}

\documentclass[twocolumn]{svjour3}

\newcommand{\comm}[1]{}

\smartqed

\usepackage{graphicx,amssymb}
\usepackage{epsf,psfig}
\usepackage{txfonts}
\usepackage{natbib}
\usepackage[unicode]{hyperref}

\journalname{Astrophysics and Space Science}

\begin{document}

\title{Escape dynamics and fractal basins boundaries in the three-dimensional Earth-Moon system}

\author{Euaggelos E. Zotos}

\institute{Department of Physics, School of Science, \\
Aristotle University of Thessaloniki, \\
GR-541 24, Thessaloniki, Greece \\
Corresponding author's email: {evzotos@physics.auth.gr}}

\date{Received: 31 December 2015 / Accepted: 31 January 2016 / Published online: 9 February 2016}

\titlerunning{Escape dynamics in the 3D Earth-Moon system}

\authorrunning{Euaggelos E. Zotos}

\maketitle

\begin{abstract}
The orbital dynamics of a spacecraft, or a comet, or an asteroid in the Earth-Moon system in a scattering region around the Moon using the three dimensional version of the circular restricted three-body problem is numerically investigated. The test particle can move in bounded orbits around the Moon or escape through the openings around the Lagrange points $L_1$ and $L_2$ or even collide with the surface of the Moon. We explore in detail the first four of the five possible Hill's regions configurations depending on the value of the Jacobi constant which is of course related with the total orbital energy. We conduct a thorough numerical analysis on the phase space mixing by classifying initial conditions of orbits in several two-dimensional types of planes and distinguishing between four types of motion: (i) ordered bounded, (ii) trapped chaotic, (iii) escaping and (iv) collisional. In particular, we locate the different basins and we relate them with the corresponding spatial distributions of the escape and collision times. Our outcomes reveal the high complexity of this planetary system. Furthermore, the numerical analysis suggests a strong dependence of the properties of the considered basins with both the total orbital energy and the initial value of the $z$ coordinate, with a remarkable presence of fractal basin boundaries along all the regimes. Our results are compared with earlier ones regarding the planar version of the Earth-Moon system.


\end{abstract}

\section{Introduction}
\label{intro}

Particles moving in escaping orbits in Hamiltonian systems is, without any doubt, one of the most interesting topics in nonlinear dynamics \citep[e.g.,][]{C90,CK92,CKK93,STN02}. Hamiltonian systems with escapes are also known as open or leaking Hamiltonian systems, in which there is a finite energy of escape. When the value of the energy is higher than the energy of escape, the equipotential surfaces open and escape channels emerge through which the test particle can escape to infinity. The literature is replete with research studies on the field of leaking Hamiltonian systems \citep[e.g.,][]{BBS09,EP14,KSCD99,LT11,NH01,Z14a,Z14b,Z15b}.

The problem of escaping orbits in open Hamiltonian systems is however less explored than the related problem of chaotic scattering. The viewpoint of chaos theory has been used in order to investigate and interpret the phenomenon of chaotic scattering \citep[e.g.,][]{BOG89,BTS96,BST98,JLS99,JMS95,JT91,SASL06,SSL07}. At this point, we would like to emphasize that all the above-mentioned references on the issues of open Hamiltonian systems and chaotic scattering are exemplary rather than exhaustive.

In leaking Hamiltonian systems an issue of great importance is the determination of the basins of escape. The basins of escape are similar to the basins of attraction in the case of dissipative systems or to the Newton-Raphson fractal basins. Many previous papers have been devoted on the study of the basins of escape \citep[e.g.,][]{BGOB88,KY91,PCOG96}. In \citep{C02} the reader can find more theoretical details regarding the basins of escape. Moreover, the escape basins in a multi-channel dynamical system of a perturbed harmonic oscillator have been studied in \citep{Z14b}. The boundaries between the several basins of escape may be fractal \citep[e.g.,][]{AVS09,BGOB88} or in the case where three or more escape channels coexist obey the more restrictive Wada property \citep[e.g.,][]{AVS01}.

The three-body problem is a paradigmatic case in celestial mechanics. This problem deals with the gravitationally interacting celestial bodies and predicts their Newtonian motion. All planets and asteroids in our Solar System move around the Sun, while in the same manner the moons orbit their host planets. The system Sun-planet-moon, or the Sun-planet-asteroid can be considered as typical examples of the three-body problem. In particular, the three-body problem of the Sun-planet-asteroid system can be significantly simplified since the mass of the asteroid is always negligible with respect to the masses of the Sun and the planet. This means that the gravitational influence of the asteroid on the Sun and also on the planet can be easily omitted from the equation describing the planetary system. Using this assumption then the three-body problem becomes the usual restricted three-body problem (RTBP) \citep{S67}. In this case there are two possibilities regrading the type of motion of the two primary bodies around their common center of mass: the circular RTBP and the elliptic RTBP.

Escaping and colliding orbits in the RTBP is another typical example. In \citet{N04} and \citet{N05} a systematic orbit classification was carried out in the planar circular RTBP. In particular, initial conditions of orbits were classified into three main categories: (i) bounded; (ii) escaping and (iii) collisional, while bounded regular orbit were further classified into orbital families regarding their motion around the two primary bodies. \citet{dAT14} (hereafter Paper I) investigated the orbital dynamics of the two dimensional version of the Earth-Moon system in a scattering region around the Moon. They explored the orbital content classifying orbits in the same categories in all possible Hill's regions configurations. In the same vein in \citet{Z15c} and \citet{Z15d} we classified orbits thus revealing the orbital structure of the Saturn-Titan and Pluto-Charon planetary systems, respectively. In the present paper we shall expand the work initiated in Paper I in order to unveil the orbital structure of the three dimensional Earth-Moon system. From dynamical astronomy point of view the 3D version of the problem is more important of the 2D due to the reality of transportation of materials throughout the planetary systems.

The structure of the paper is as follows: In Section \ref{mod} we present in detail the properties of the mathematical model. All the computational methods we used in order to obtain the classification of the orbits are described in Section \ref{cometh}. In the following Section, we conduct a thorough and systematic numerical investigation revealing the overall orbital structure of the three dimensional Earth-Moon planetary system by classifying orbits into categories. Our paper ends with Section \ref{disc} where the discussion and the conclusions of our research are given.

\section{Presentation of the mathematical model}
\label{mod}

At this point, it would be beneficial to recall the basic properties of the mathematical model which describes the circular restricted three-body problem (CRTBP) \citep{S67}. The motion of both primary bodies $P_1$ and $P_2$ is circular around their common center of mass, while the test particle (which is the third body) moves under the gravitational filed created by the two primaries. Taking into account that the mass of the test particle is significantly smaller than the masses of the primaries we may reasonably assume that the third body does not influence or perturb the motion of the primaries. The mass ratio $\mu = m_2/(m_1 + m_2)$, $m_1 > m_2$ is used to define the non-dimensional masses of the primaries as $1-\mu$ and $\mu$. For the Earth-Moon planetary system we have $\mu$ = 0.0121506683\footnote{For the value of the mass ratio $\mu$ we adopted the same precision (number of significant decimal digits) as in Paper I.}. The centers of the primaries are both located on the $x$-axis at $(-\mu, 0, 0)$ and $(1-\mu, 0, 0)$. For the description of the motion of the test particle we choose a rotating coordinate frame of reference where its origin is at the center of mass of the two primaries.

The total time-independent effective potential function in the rotating frame of reference is
\begin{equation}
\Omega(x,y,z) = \frac{(1 - \mu)}{r_1} + \frac{\mu}{r_2} + \frac{\mu \left(1 + \mu \right)}{2} + \frac{1}{2}\left( x^2  + y^2 \right),
\label{pot}
\end{equation}
where
\[
r_1 = \sqrt{\left(x + \mu\right)^2 + y^2 + z^2},
\]
\begin{equation}
r_2 = \sqrt{\left(x + \mu - 1\right)^2 + y^2 + z^2},
\label{dist}
\end{equation}
are the distances to the respective primaries. Here we would like to point out that we expanded into three dimensions the effective potential function $\Omega(x,y,z)$ used in Paper I, where the third term (which is not present in the classical CRTBP) is only a constant which controls the critical values of the Jacobi constant but it does not affect the motion of the test particle.

The motion of the test particle in this corotating frame of reference is described by the following scaled equations of motion
\[
\Omega_x = \frac{\partial \Omega}{\partial x} = \ddot{x} - 2\dot{y},
\]
\[
\Omega_y = \frac{\partial \Omega}{\partial y} = \ddot{y} + 2\dot{x},
\]
\begin{equation}
\Omega_z = \frac{\partial \Omega}{\partial z} = \ddot{z}.
\label{eqmot}
\end{equation}

It is well known that the dynamical system (\ref{eqmot}) admits only one integral of motion which is known as the Jacobi integral of motion.
\begin{equation}
J(x,y,z,\dot{x},\dot{y},\dot{z}) = 2\Omega(x,y,z) - \left(\dot{x}^2 + \dot{y}^2 + \dot{z}^2 \right) = C.
\label{ham}
\end{equation}

A five-dimensional (5D) invariant manifold inside the six-dimensional (6D) phase space is defined through the Jacobi constant $C$. This means that for a given value of the total orbital energy the motion of the test particle is restricted in the domain where $C \leq 2 \Omega(x,y,z)$, while all the other areas are energetically forbidden. The Jacobi constant $C$ and the total orbital energy $E$ are connected through the relation $C = - 2E$.

Lagrange found that for bodies which move in circular orbits around their common center of gravity there are five distinct three-body formations. These formations appear to be invariant for an observer in the rotating frame of reference. Moreover, these five locations of the test particle for which its location appears to be stationary or invariant when viewed from the rotating frame are called Lagrange libration points $L_i$, $i = 1, ..., 5$ at which
\begin{equation}
\frac{\partial \Omega}{\partial x} = \frac{\partial \Omega}{\partial y} = \frac{\partial \Omega}{\partial z} = 0.
\label{lps}
\end{equation}
$L_1$, $L_2$, and $L_3$ (known as collinear points) are located on the $x$-axis, while the other two $L_4$ and $L_5$ are called triangular points and they are located in the vertices of equilateral triangles. It should be noted that the labeling of the first three Lagrange points is not consistent throughout the literature. In our case, we use the most popular labeling according to which $L_1$ is between the two primaries, $L_2$ is at the right side of $P_2$, while $L_3$ is at the left side of $P_1$. The values of the Jacobi integral of motion at the Lagrange points $L_i, i = 1, ..., 5$ are denoted by $C_i$. For the Earth-Moon system we have: $C_1 \approx 3.20034491$, $C_2 \approx 3.18416414$, $C_3 \approx 3.02415026$, and $C_4 = C_5$ = 3. Once more, the accuracy of $C_i$ is the same as in Paper I.

\begin{figure*}
\centering
\resizebox{\hsize}{!}{\includegraphics{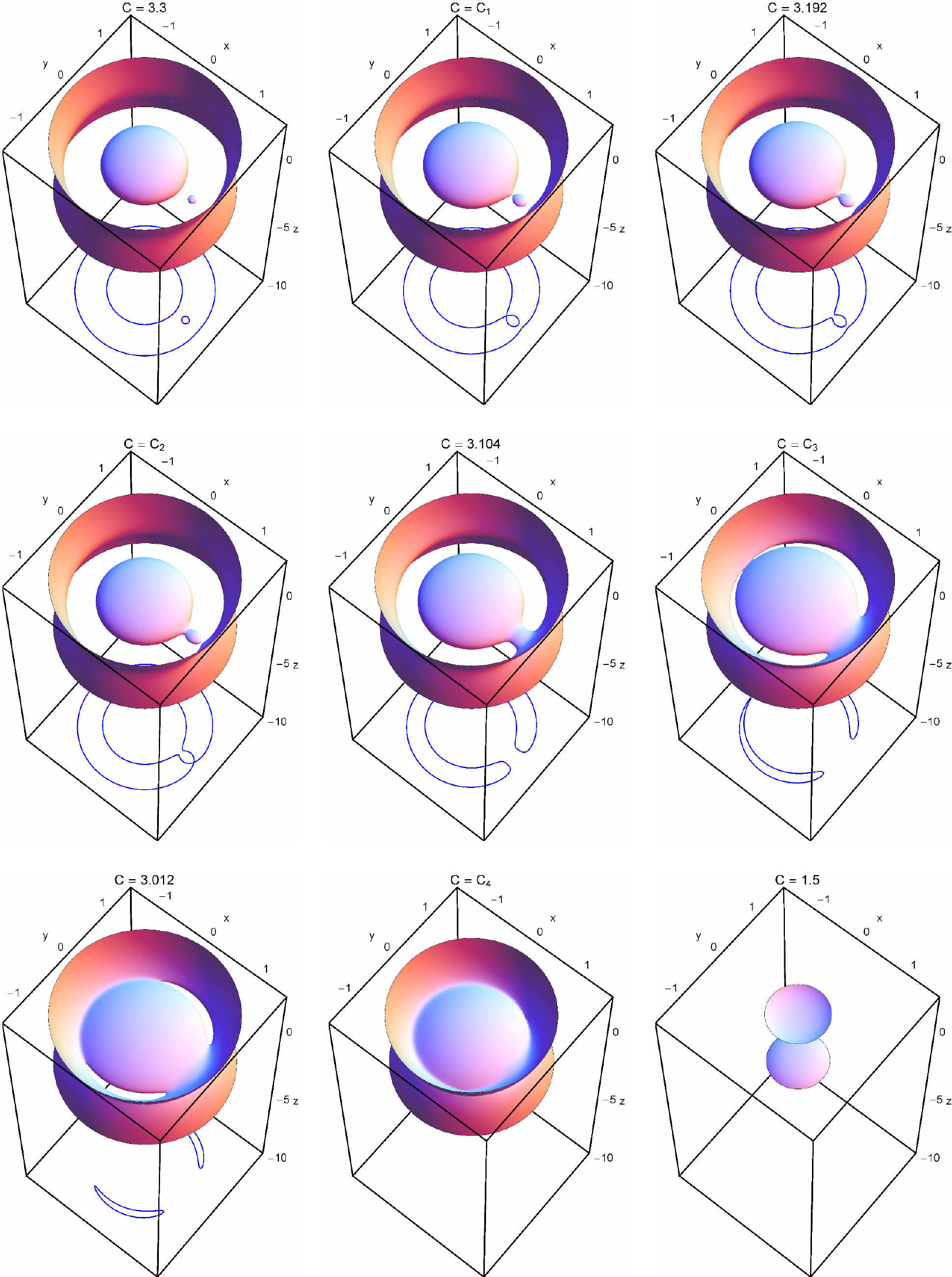}}
\caption{Evolution of the structure of three-dimensional zero velocity surfaces for the CRTBP as a function of the Jacobi constant $C$. The projections of these surfaces onto the configuration $(x,y)$ plane are depicted at the bottom of the bounding box as blue solid lines.}
\label{hrcs}
\end{figure*}

The three-dimensional (3D) surfaces described by the relation $2\Omega(x,y,z) = C$, also known as zero velocity surfaces, determine the geometric extremes possible for a given value of the energy. These 3D surfaces allow us to determine if it is possible for the test particle to travel from Earth to Moon or beyond. The projection of these surfaces onto the configuration (or position) space $(x,y)$ is called the Hill's regions. Moreover the boundaries of the Hill's regions are called zero velocity curves (ZVCs) because they are the locus in the configuration $(x,y)$ space where the kinetic energy vanishes. The structure of the zero velocity surfaces and the corresponding Hill's regions strongly depends on the value of the Jacobi constant. In particular, there are five distinct energy regions regarding the Hill's regions configurations:
\begin{itemize}
  \item Energy region I: $C > C_1$: All transport channels are closed, so there are only bounded and collision basins.
  \item Energy region II: $C_1 > C > C_1$: Only the channel around $L_1$ is open thus allowing orbits to enter the Earth realm.
  \item Energy region III: $C_2 > C > C_3$: The channel around $L_2$ is open, so orbits can enter the exterior region and escape form the system.
  \item Energy region IV: $C_3 > C > C_4$: Both channels around $L_2$ and $L_3$ are open, therefore orbits are free to escape through two different directions.
  \item Energy region V: $C < C_4$: The forbidden regions disappear, so motion over the entire configuration space is possible.
\end{itemize}
In Fig. \ref{hrcs} we present the evolution of the structure of the 3D zero velocity surfaces for several values of the Jacobi constant. It is evident that as the value of the Jacobi constant decreases several doorways appear through which the test particle can enter the several allowed region of motion.

\section{Computational methods and criteria}
\label{cometh}

For the numerical investigation of the orbital structure of the 3D Earth-Moon system we have to integrate sets of initial conditions of orbits thus classifying them into categories. In systems with three degrees of freedom however, the phase space is six-dimensional and thus the behavior of the orbits cannot be easily visualized. One easy way to overcome this issue is to work in phase spaces with lower dimensions. Let us start with initial conditions on a two-dimensional plane following the approach used successfully in \citet{ZCar13} and \citet{Z14a,Z15a}. Following this method we are able to classify initial conditions of orbits and visualize our results, if we restrict our exploration to a subspace of the whole 6D phase space. In fact we choose a value of $z_0$ which defines a two-dimensional $(x,y)$ slice inside the three-dimensional zero velocity surface. For a specific value of $z_0$ we consider dense, uniform grids of $1024 \times 1024$ initial conditions $(x_0, y_0)$ regularly distributed on the configuration $(x,y)$ plane inside the area allowed by the corresponding value of the energy. Following a typical approach, all orbits are launched with initial conditions inside a certain region, called scattering region, which in our case is $x_{L_1} \leq x \leq x_{L_2}$ and $-0.2 \leq y \leq 0.2$. For all 3D orbits $\dot{x_0} = \dot{z_0} = 0$, while the value of $\dot{y_0}$ is always obtained from the Jacobi integral of motion (\ref{ham}) as $\dot{y_0} = \dot{y}(x_0,y_0,z_0,\dot{x_0},\dot{z_0},C) > 0$\footnote{In Paper I the authors classified two dimensional orbits with $\dot{x_0} = 0$ and $\dot{y_0} > 0$. In our work we expand into three dimensions this choice of initial conditions thus considering orbits with $\dot{x_0} = \dot{z_0} = 0$ and $\dot{y_0} > 0$. The case with non zero values in $x$ and $z$ components of the velocity of the test particle is another choice which however is not considered here, obviously for saving space.}.

The value of the Jacobi constant defines the structure of the zero velocity surfaces. In particular, the region around the Moon can be connected or not to the Earth realm through the neck around $L_1$ or to the exterior region through the opening around $L_2$. Therefore the initial conditions of orbits inside the scattering region can be classified into categories: (i) bounded orbits which remain in the Moon realm, (ii) escaping orbits to the Earth realm through $L_1$, (iii) escaping orbits to the exterior realm through $L_2$, and (iv) orbits that collide with the Moon.

Appropriate numerical criteria are needed in order to distinguish between the four types of motion. Following the approach used in Paper I, we assume that the Moon is a finite body with a mean radius of approximately 1738 km (about $4.521 \times 10^{-3}$ dimensionless length units). Therefore, if an orbit reaches the surface of the Moon its numerical integration ends thus producing an orbit leaking in the phase space. For escaping orbits we shall extend in three dimensions the escape criteria used in Paper I. In particular, an escaping orbit to the Earth realm must satisfy the conditions $x < x_{L_1} - \delta_1$, with $\delta_1 = 0.15$ and the third body inside the sphere around the the Moon of radius $r_1 \leq |x_{L_3} - x_{P_1}|$. In the same vein, an escaping orbit to the exterior realm must fulfill the conditions $x > x_{L_2} + \delta_2$, with $\delta_2 = 0.09$, or $|y| > y_{L_5}$, or, if $x < x_{L_1} - \delta_1$, $r_1 > |x_{L_3} - x_{P_1}|$ (i.e., the third body is outside the sphere just defined). It should be clarified that the tolerances $\delta_1$ and $\delta_2$ were included in the escape criteria so as to avoid the unstable Lyapunov orbits \citep{L49} to be incorrectly classified as escaping orbits.

In the CRTBP the phase space is divided into the collision, escaping and non-escaping regimes. In most cases, the non-escaping regions correspond to ordered orbits forming stability islands. In many Hamiltonian systems however, trapped chaos have also been observed \citep[e.g.,][]{Z15a}. Therefore, we decided to distinguish between trapped chaotic motion and non-escaping ordered motion. A plethora of chaos indicators is available for distinguishing between chaotic and ordered motion. In this study we choose the Smaller ALingment Index (SALI) which has been proved, over the years, a very reliable, fast and effective tool \citep[e.g,][]{S01}.

The maximum time $t_f$ of the numerical integration was set to be equal to $10^4$ dtu (dimensionless time units), corresponding to about 128.63 yr. Needless to say that when an orbit collided with the Moon or escaped the numerical integration was effectively ended. Our choice of such a long integration time can be justified by the presence of ``sticky orbits" which behave for long time intervals as regular ones before they exhibit their true chaotic nature. So we wanted to be sure that all initial conditions of orbits have enough integration time for revealing their character.

A double precision Bulirsch-Stoer \verb!FORTRAN 77! algorithm \citep[see e.g.,][]{PTVF92} was used in order to forward integrate the equations of motion (\ref{eqmot}) as well as the variational equations for all the initial conditions of the orbits. The error regarding the energy conservation throughout our computations was smaller than $10^{-12}$, which is sufficiently low. For initial conditions leading to collision with the Moon, where the test particle moves inside a region of radius $10^{-2}$ around the same primary, we applied the Lemaitre's global regularization method \citep[e.g.,][]{S67}. All graphics presented in this work have been created using version 10.3 of the software Mathematica$^{\circledR}$ \citep{W03}.

\section{Numerical results - Orbit classification}
\label{numres}

This section is devoted on the classification of orbits in the configuration $(x,y)$ space into five categories: (i) bounded regular orbits; (ii) trapped chaotic orbits; (iii) escaping orbits to the Earth realm; (iv) escaping orbits to the exterior realm and (v) collisional orbits. Furthermore, two additional dynamical quantities of the orbits will be examined: (i) the time-scale of the collision and (ii) the time-scale of the escape. Our main numerical task will be to explore these properties for various values of $z_0$ and of the total orbital energy, or in other words for various values of the Jacobi constant. In particular, we will consider four different energy cases which correspond to the first four possible Hill's regions configurations. In the last Hill's regions configuration the escape channels lose their meaning due to the lack of forbidden regions, so we cannot distinguish between escape to the Earth realm and to the exterior realm.

In the following color-coded orbit type diagrams (OTDs) each pixel is assigned a color according to the type of the orbit. Therefore, all initial conditions of orbits can be classified into bounded orbits, unbounded or escaping orbits and collisional orbits. In this special type of Poincar\'{e} Surface of Section (PSS) the phase space emerges as a close and compact mix of escape basins\footnote{The set of initial conditions of orbits which lead to a certain final state (escape, collision or bounded motion) is defined as a basin.}, collision basins and stability regions. Our initial numerical computations suggest that a non-zero amount of non-escaping orbits is present along with escaping and collisional orbits. Generally speaking, most of the non-escaping basins are composed of initial conditions of ordered orbits, where an adelphic integral of motion is present. This additional integral of motion restricts their accessible phase space and therefore it hinders their escape from the system.

\begin{figure*}
\centering
\resizebox{\hsize}{!}{\includegraphics{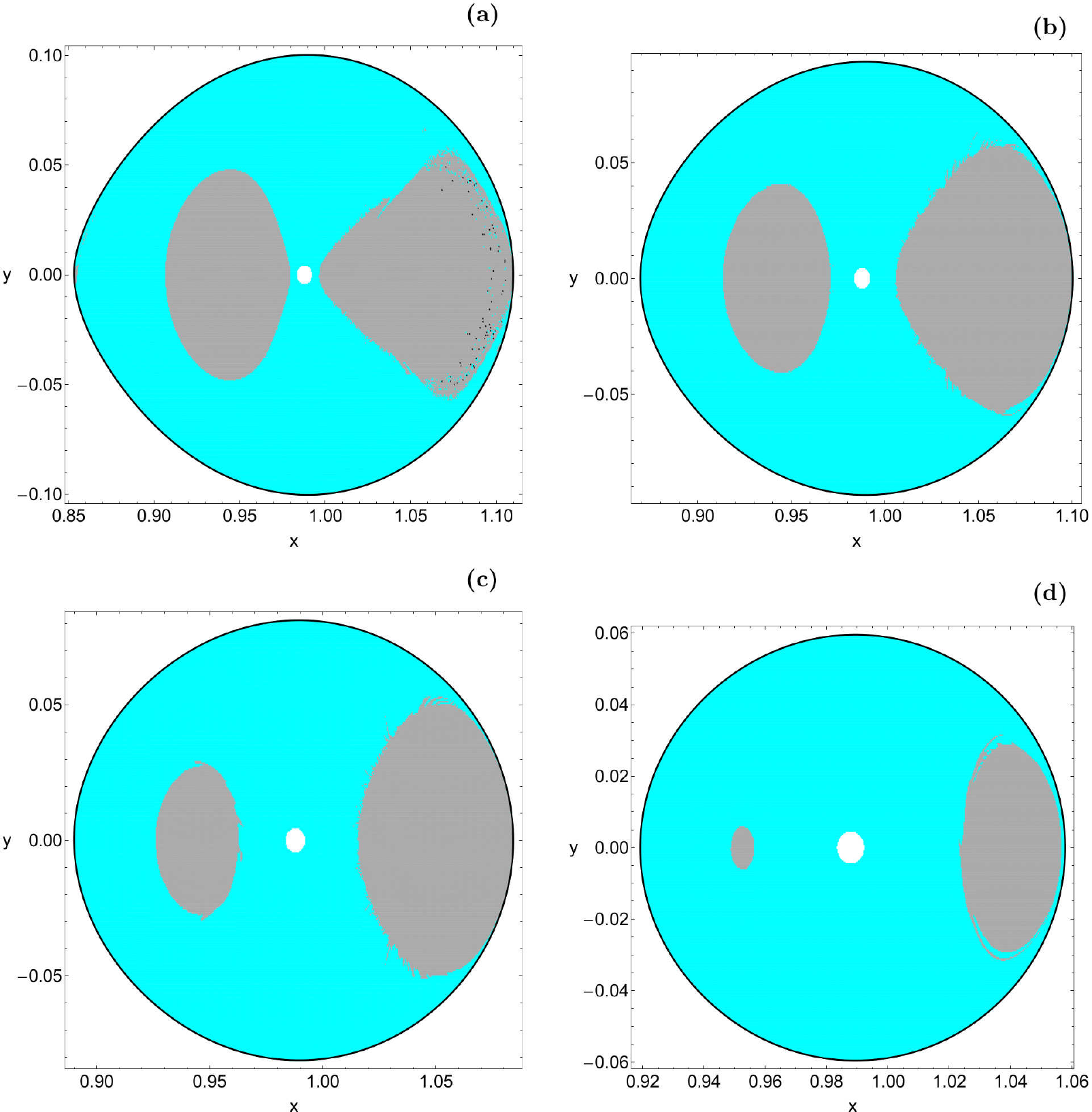}}
\caption{Basin diagrams for energy region I when $C = 3.201$. (a-upper left): $z_0 = 0.02$, (b-upper right): $z_0 = 0.04$, (c-lower left): $z_0 = 0.06$ and (d-lower right): $z_0 = 0.08$. The color code is as follows: bounded basins of regular orbits (gray), trapped chaotic orbits (black) and collision basins (cyan).}
\label{hr1}
\end{figure*}

\begin{figure*}
\centering
\resizebox{\hsize}{!}{\includegraphics{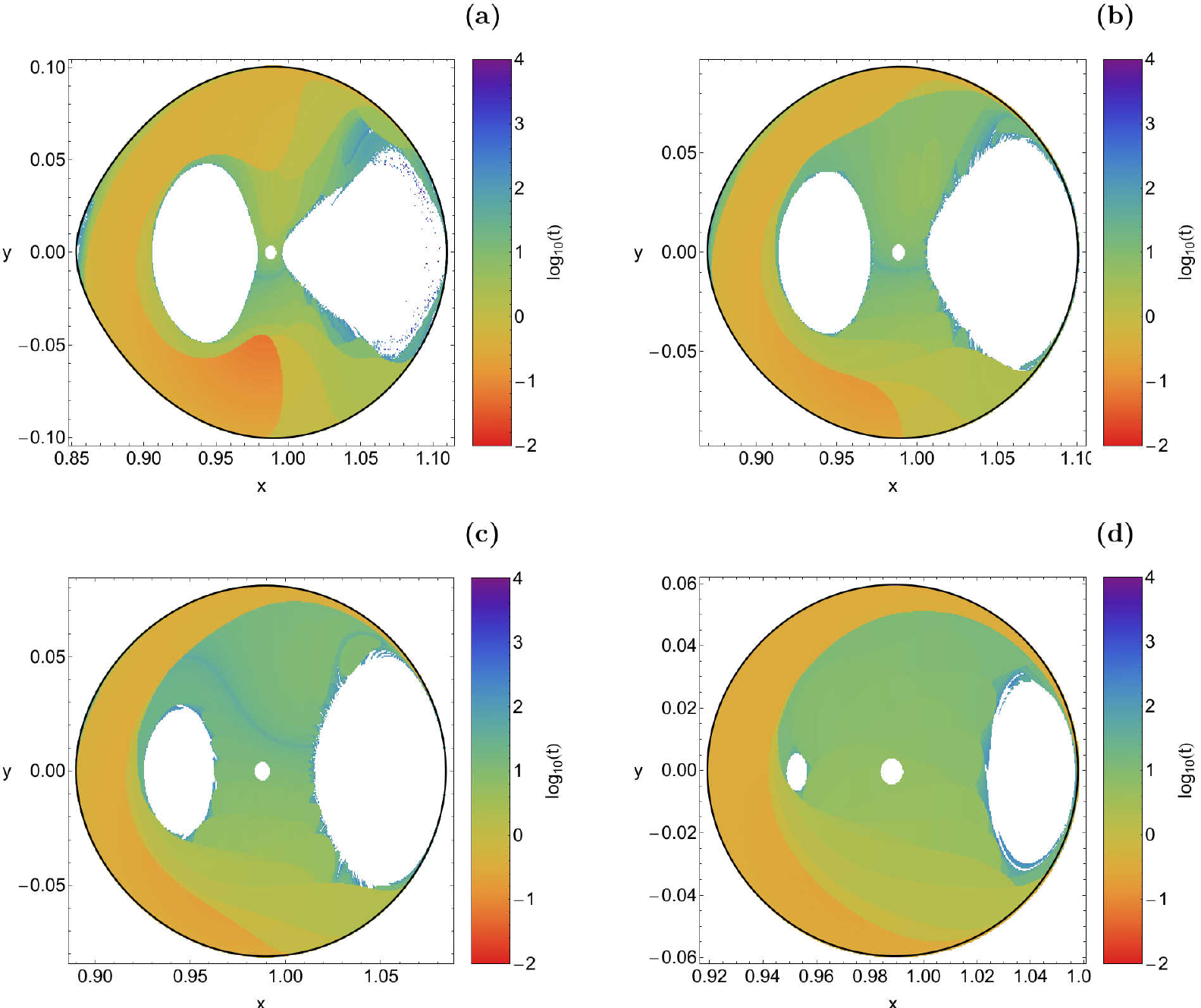}}
\caption{Distribution of the collision time of the orbits on the configuration $(x,y)$ space for the values of $z_0$ of Fig. \ref{hr1}(a-d). The bluer the color, the larger the collision time. Initial conditions of bounded regular orbits and trapped chaotic orbits are shown in white.}
\label{hr1t}
\end{figure*}

\begin{figure*}
\centering
\resizebox{\hsize}{!}{\includegraphics{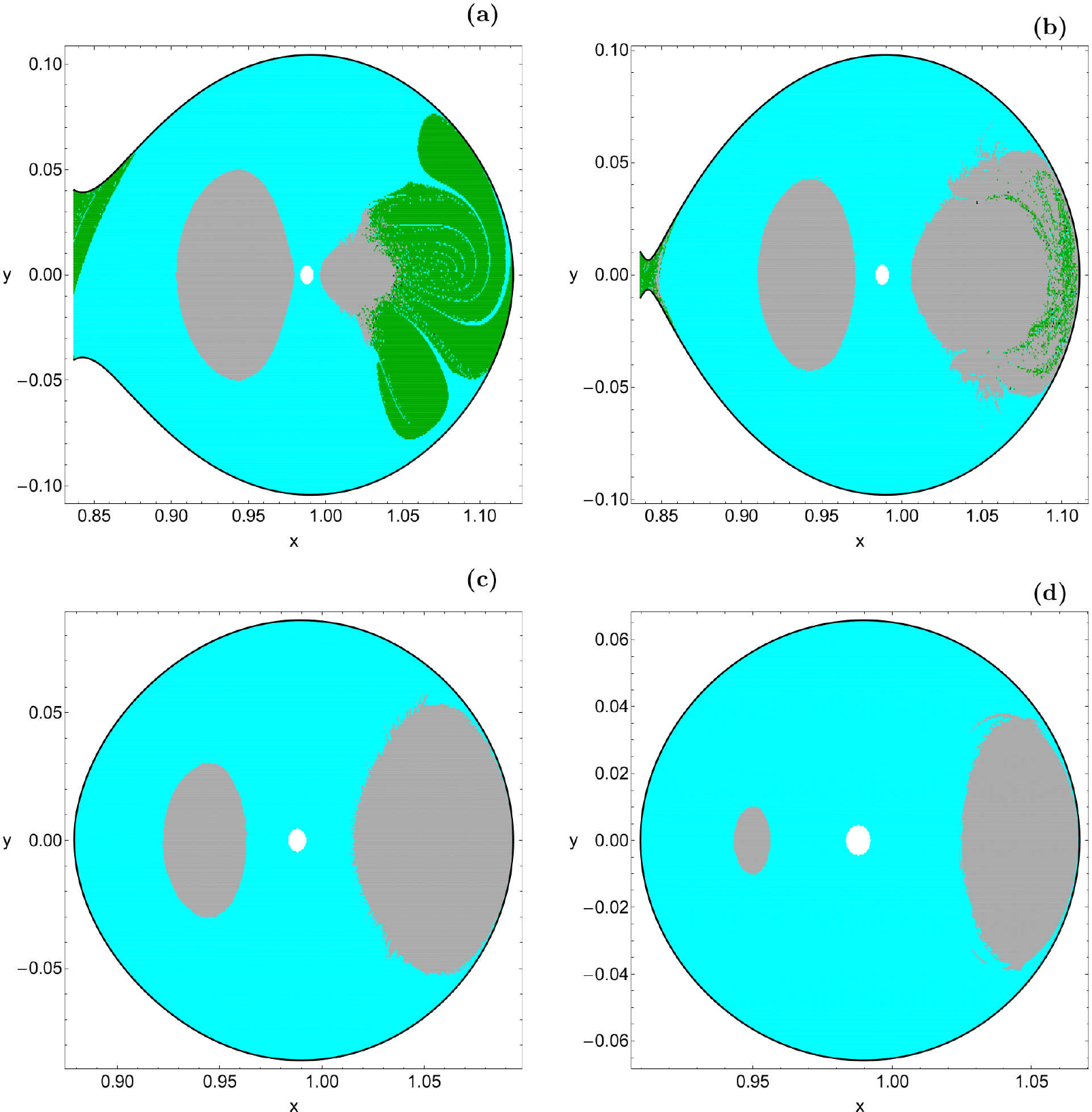}}
\caption{Basin diagrams for energy region II when $C = 3.192$. (a-upper left): $z_0 = 0.02$, (b-upper right): $z_0 = 0.04$, (c-lower left): $z_0 = 0,06$ and (d-lower right): $z_0 = 0.08$. The color code is as follows: bounded basins (gray), trapped chaotic orbits (black), collision basins (cyan), Earth realm basins (green).}
\label{hr2}
\end{figure*}

\begin{figure*}
\centering
\resizebox{\hsize}{!}{\includegraphics{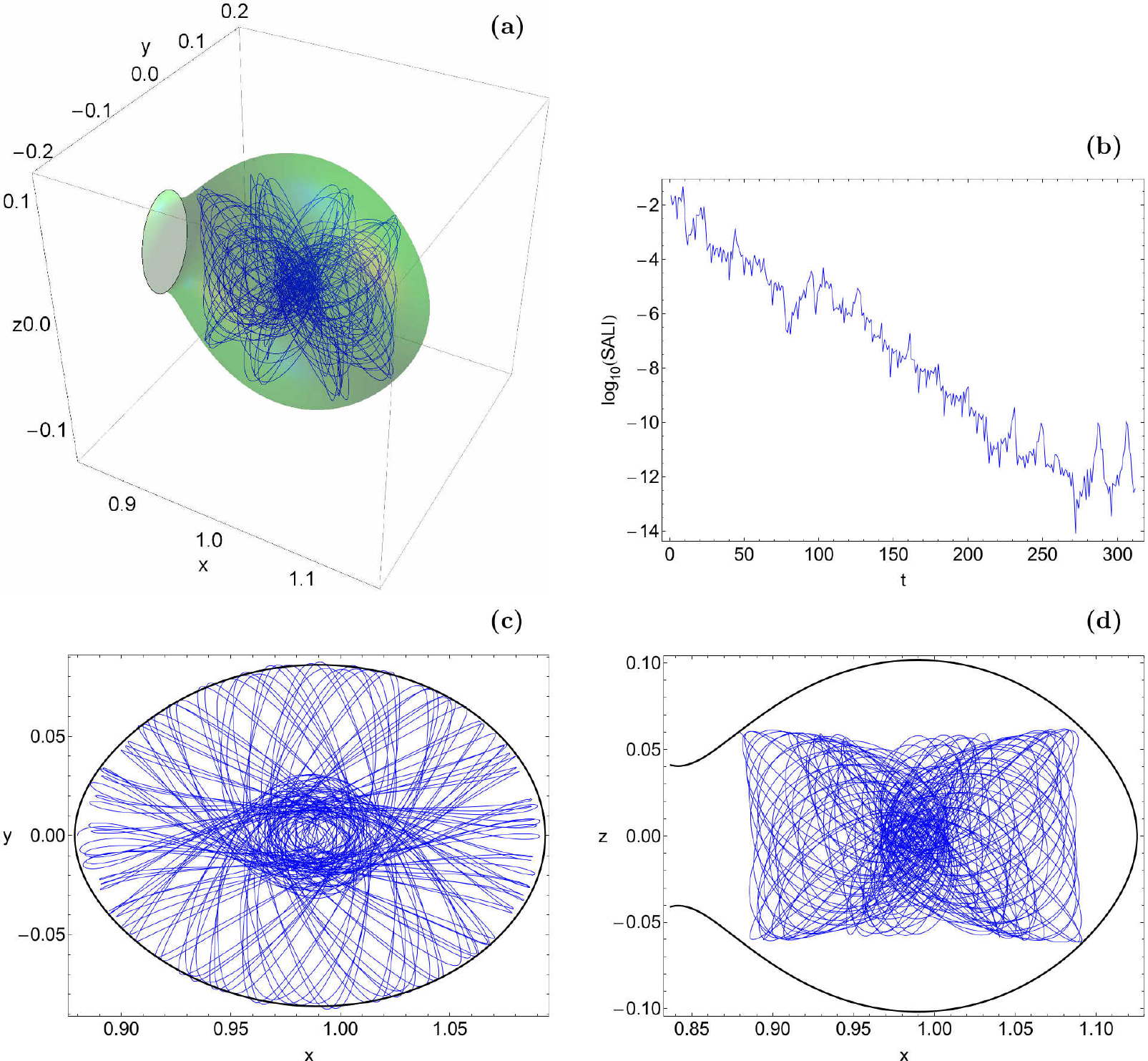}}
\caption{(a-upper left): A 3D trapped chaotic orbit in the Moon realm which collides with the Moon after about 312 dimensionless time units, (b-upper right): The time-evolution of SALI and (c-d): The 2D projections of the 3D orbit on the $(x,y)$ and $(x,z)$ planes, respectively.}
\label{orb}
\end{figure*}

\begin{figure*}
\centering
\resizebox{\hsize}{!}{\includegraphics{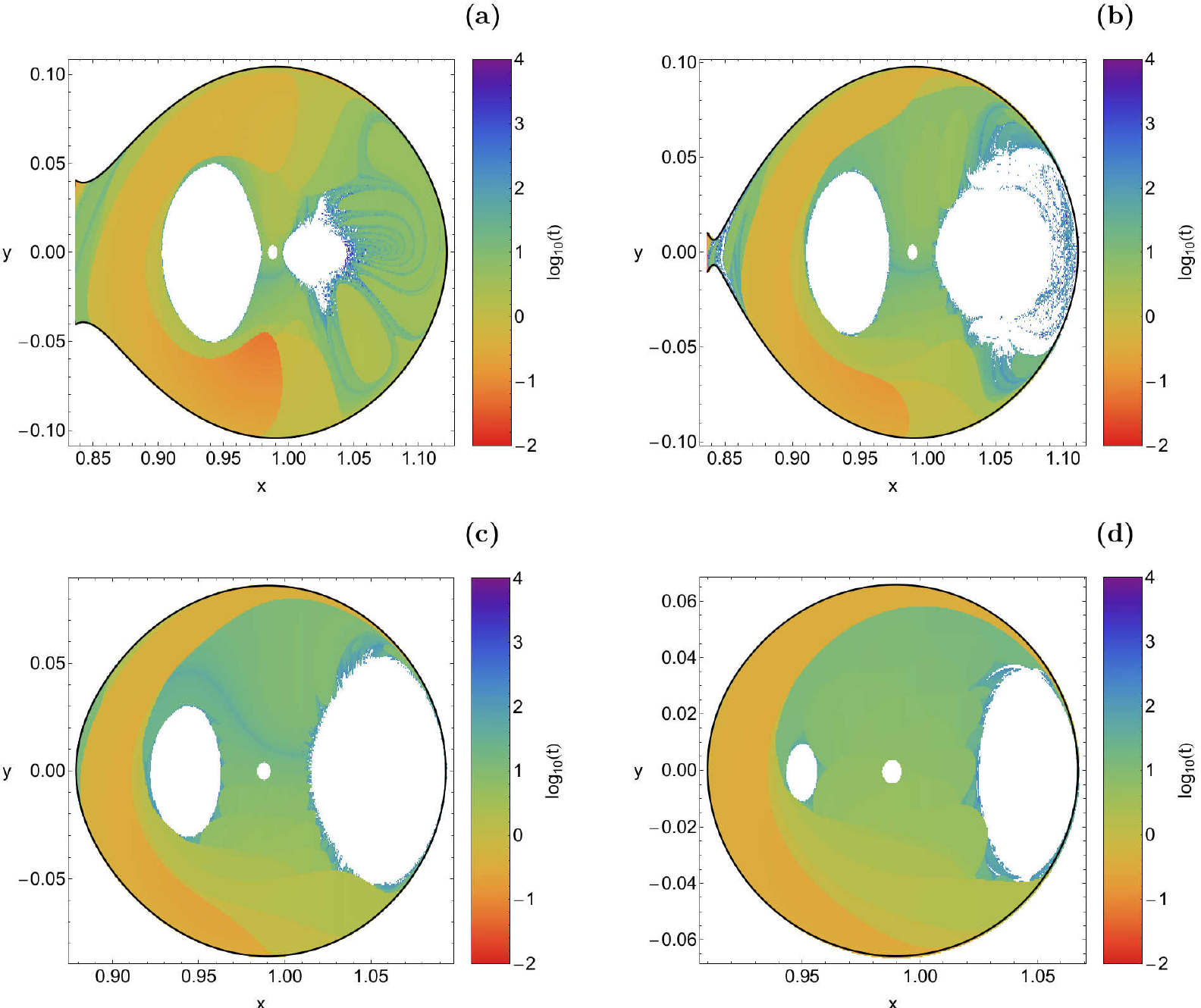}}
\caption{Distribution of the collision and escape time of the orbits on the configuration $(x,y)$ space for the values of $z_0$ of Fig. \ref{hr2}(a-d).}
\label{hr2t}
\end{figure*}

\begin{figure*}
\centering
\resizebox{\hsize}{!}{\includegraphics{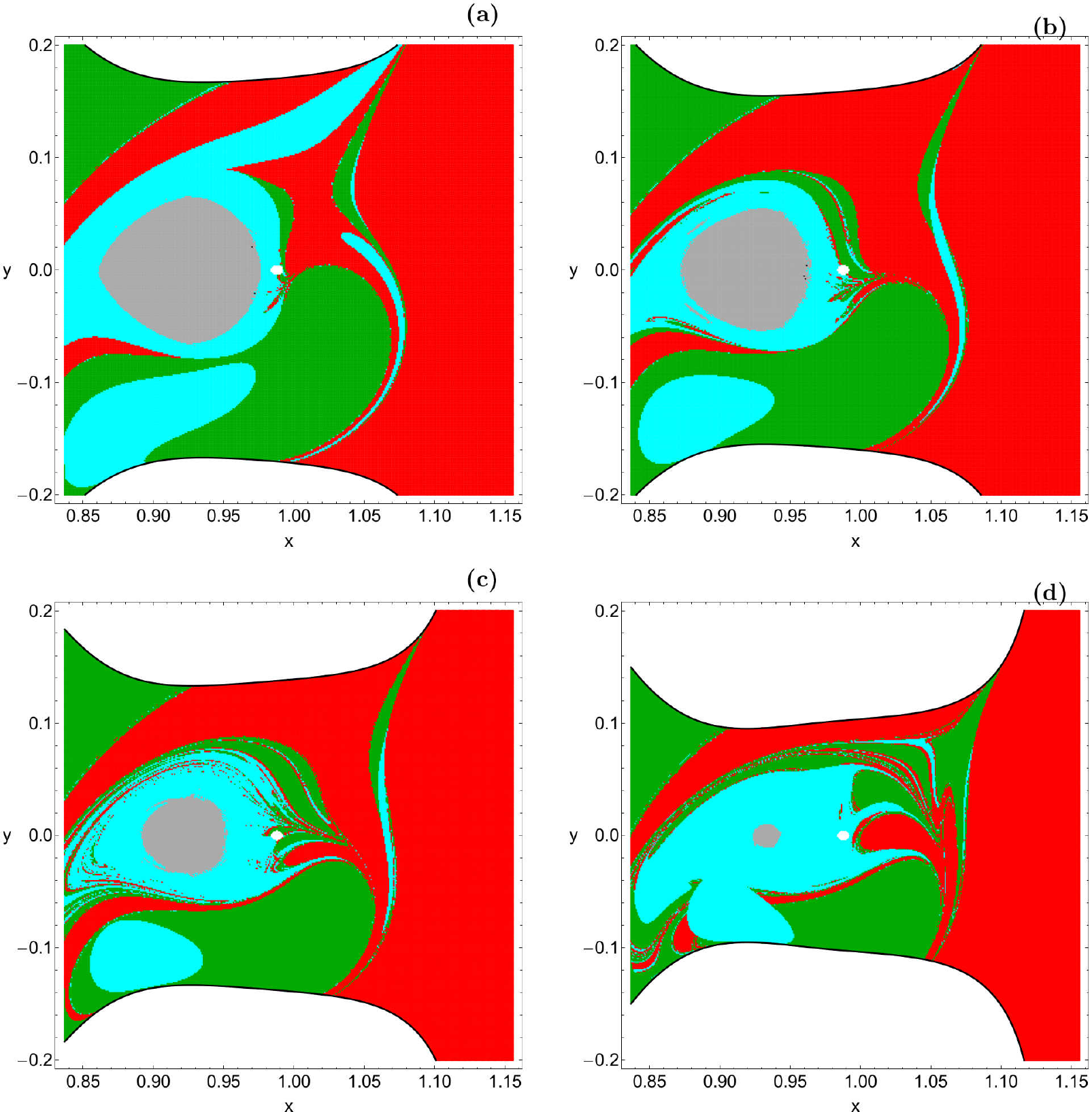}}
\caption{Basin diagrams for energy region III when $C = 3.104$. (a-upper left): $z_0 = 0.03$, (b-upper right): $z_0 = 0.06$, (c-lower left): $z_0 = 0.09$ and (d-lower right): $z_0 = 0.12$. The color code is as follows: bounded basins (gray), trapped chaotic orbits (black), collision basins (cyan), Earth realm basins (green), exterior realm basins (red).}
\label{hr3}
\end{figure*}

\begin{figure*}
\centering
\resizebox{\hsize}{!}{\includegraphics{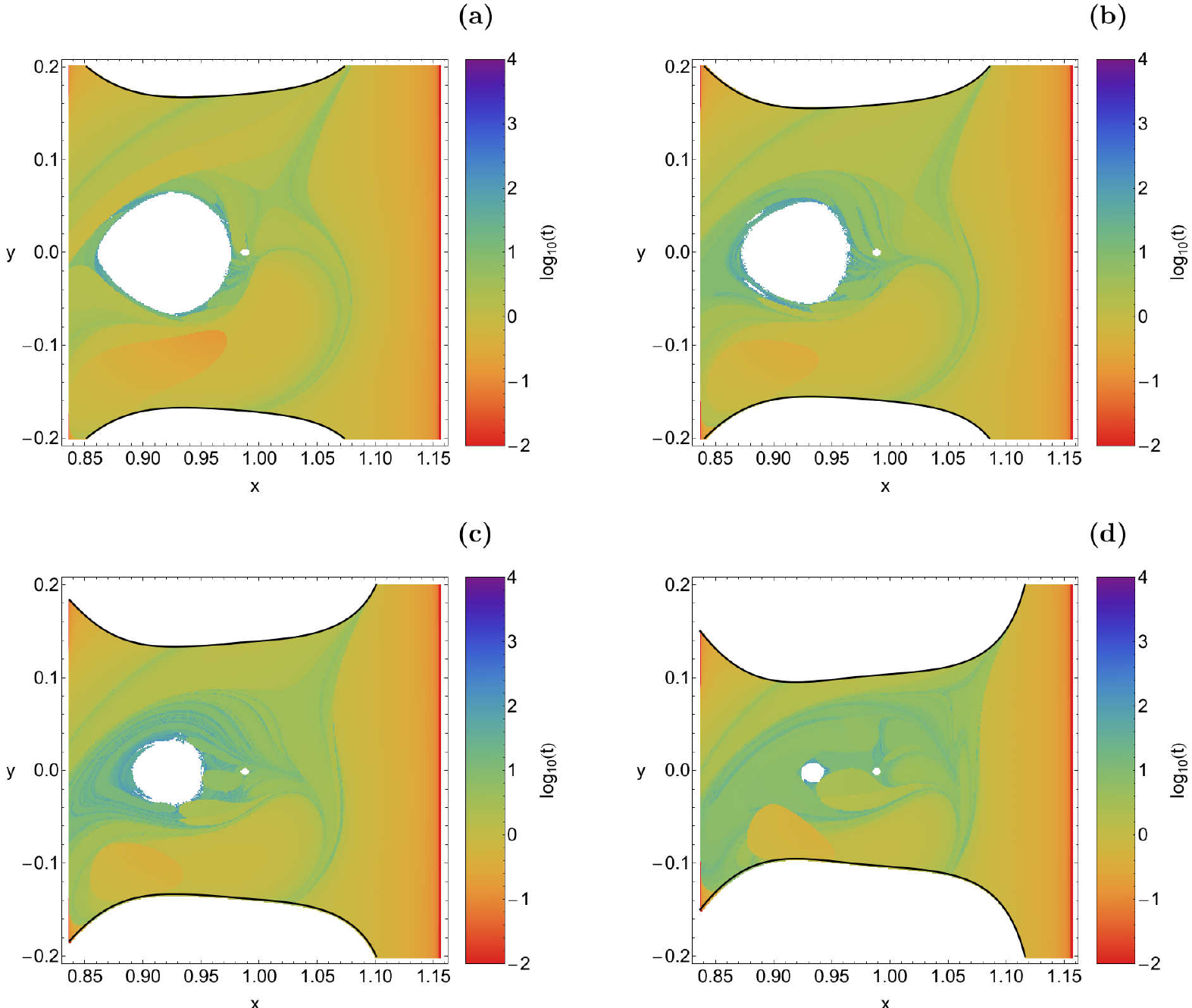}}
\caption{Distribution of the collision and escape time of the orbits on the configuration $(x,y)$ space for the values of $z_0$ of Fig. \ref{hr3}(a-d).}
\label{hr3t}
\end{figure*}

\subsection{Energy region I: $C > C_1$}
\label{ss1}

The color-coded OTDs for four values of $z_0$ when $C = 3.201$ are presented in Fig. \ref{hr1}(a-d). The outermost black solid line is the ZVC on the configuration $(x,y)$ space which is defined as $2\Omega(x,y) = C$. Inspecting these diagrams we observe that in energy region I there is only bounded and collisional motion. In particular, for $z_0 = 0.02$, we see in Fig. \ref{hr1}a that two stability islands are present inside the unified collision basin. In contrast to Paper I we decided to exclude initial conditions of orbits inside the radius of the Moon because these orbits lack of physical meaning. Therefore, the white hole around the center of the Moon shown in Fig. \ref{hr1}a represents an additional ``forbidden" region\footnote{Obviously if we numerically integrate initial conditions inside this region we will see that they lead to immediate collision to the Moon.}. The stability island situated on the right side of the Moon, surrounds a retrograde (clockwise) symmetric periodic orbit around $P_2$. On the other hand, the stability island, located on the left side of the Moon, surrounds a periodic orbit around $P_2$ which is symmetric to a reflection over the $x$-axis $(y = 0)$ and is traveled in counterclockwise sense, hence, prograde with respect to the rotating coordinate system. In the case of the planar Hill problem it was proved that the stability island located on the left side of the Moon is much more stable than that on the right side in relation to energy variation \citep{SS00}. It is worth noticing that at the boundaries of the stability island on the right side of the Moon we identified several individual initial conditions corresponding to trapped chaotic orbits (shown as black dots). As the value of $z_0$ increases the area of the ZVC is reduced thus leading to motion close to the Moon and therefore the area of the central white forbidden region increases. Moreover it is evident in Figs. \ref{hr1}(b-d) that the size of both stability islands reduces with increasing $z_0$. Additional numerical experiments (not shown here) using other values of $C$, always for $C > C_1$, suggest that in energy region I collision motion dominates for orbits with relative high values of $z_0$.

In the following Fig. \ref{hr1t}(a-d) we show how the collision times of orbits are distributed on the configuration $(x,y)$ space for the four values of $z_0$ discussed in Fig. \ref{hr1}(a-d). Light reddish colors correspond to fast collisional orbits, dark blue/purple colors indicate large collision times, while white color denote stability islands of regular motion and trapped chaotic motion. Note that the scale on the color bar is logarithmic. Looking carefully at Fig. \ref{hr1t}(a-d) we clearly observe that in the area between the two stability islands the collision times of orbits are higher than those of the initial conditions located in the left side of the plane.

\subsection{Energy region II: $C_1 > C > C_2$}
\label{ss2}

When $C < C_1$ the neck around $L_1$ opens thus allowing orbits to enter the Earth realm. Indeed for $C = 3.192$ and $z_0 = 0.02$ we see in Fig. \ref{hr2}a that there is an escape channel in the left side of the ZVC. Near the Lagrange point $L_1$ we observe an escape basin, while at the boundaries of the right stability region we see a highly-fractal\footnote{When we state that an area is fractal we simply mean that it has a fractal-like geometry without conducting any specific calculations as in \citet{AVS09}.} structure by constructed by initial conditions of escaping orbits. For $z_0 = 0.04$ the width of the escape channel reduces as it is seen in Fig. \ref{hr2}b. Now the fractal structure at the right stability island has been significantly reduced, while the size of the stability island has grown. As the value of $z_0$ increases the width of the escape channel decreases as we proceed to higher $z_0$ slices of the three-dimensional energy surface. In Fig. \ref{hr2}c where $z_0 = 0.06$ one can see that the ZVC is closed. This however does not mean that the transport channel to the Earth realm is closed. It simply means that the $z_0 = 0.06$ slice is above the escape channel. Here the orbital structure of the configuration space is very similar to that discussed earlier in energy region I. Indeed, there is no evidence of escape motion to Earth realm whatsoever. On the contrary the entire $(x,y)$ plane is covered by either collisional orbits or non-escaping regular orbits. The same pattern continues in Fig. \ref{hr2}d where $z_0 = 0.08$. It is seen that the extent of both stability islands has been reduced, while all fractal areas have disappeared since the basin boundaries are very smooth.

We numerically integrated additional sets of orbits for other values of the Jacobi constant in the same energy region $(C_1 > C > C_2)$ and for several values of $z_0$. In all examined cases we observed that for $z_0$ slices above the escape channel the test particle does not enter the Earth realm, even though the corresponding escape channel is wide open. In Fig. \ref{orb}(a-d) we provide a characteristic example of a 3D orbit with initial conditions: $x_0 = 0.88, y_0 = 0, z_0 = 0.06, \dot{x_0} = \dot{z_0} = 0$, while $\dot{y_0}$ is obtained from the Jacobi integral of motion for $C = 3.192$. In Fig. \ref{orb}a we see the orbit inside the three-dimensional energy shell, while in panels (c) and (d) we present the two-dimensional projections of the orbit on the $(x,y)$ and $(x,z)$ planes, respectively. The time-evolution of SALI in panel (b) clearly indicates that the orbit is chaotic. In fact this orbit moves chaotically around the Moon for about 312 time units and then it collides with the Moon. The interesting phenomenon is that the test particle does not enter the Earth realm even though the transport channel is open. Our numerical analysis suggests that 3D orbits with $z_0$ higher than the height of the escape channel remain trapped in the Moon realm and eventually collide with it.

At this point we would like to emphasize that the numerical integration is stopped as soon as the test particle passes $L_1$ thus entering the Earth realm. However, if we do not stop the numerical integration some orbits that initially entered the Earth realm may return to the Moon's region, or collide into one of the primary bodies, or even enter the exterior region and then escape to infinity. In our calculations however, we follow the approach used in Paper I and we consider an orbit to escape to Earth realm if the test particle passes $L_1$ even if its true asymptotic behaviour at very long time limit is different.

The distribution of the escape and collision times of orbits on the configuration $(x,y)$ space is shown in Fig. \ref{hr2t}(a-d). One may observe that the results are very similar to those presented earlier in Fig. \ref{hr1t}(a-d), where we found that orbits with initial conditions inside the escape and collision basins have the smallest escape/collision rates, while on the other hand, the longest escape/collision rates correspond to orbits with initial conditions in the fractal regions of the OTDs (see e.g., panels (a) and (b) of Fig. \ref{hr2t}). Inspecting the spatial distribution of various different ranges of escape time, we are able to associate medium escape time with the stable manifold of a non-attracting chaotic invariant set, which is spread out throughout this region of the chaotic sea, while the largest escape time values on the other hand, are linked with sticky motion around the stability islands.

\subsection{Energy region III: $C_2 > C > C_3$}
\label{ss3}

This energy region constitutes the Hill's regions with the most interest in point of view of planetary systems. Furthermore, it has many practical applications, such as orbit determination of spacecraft mainly based on many-models and also the phenomenon of temporary capture of comets or asteroids around planets in our solar system. In this energy region the exterior realm basins arise competing for the chaotic sea with the Earth realm basins and the collision set. The exterior realm rises when $C < C_2$. In Fig. \ref{hr3}(a-d) we see that for $C = 3.104$ both necks around $L_1$ and $L_2$ are open. In Fig. \ref{hr3}a where $z_0 = 0.03$ we observe that the configuration $(x,y)$ space is covered by all types of basins. At the left side of the Moon collision basins and Earth realm basins dominate, while at the right side of the Moon the majority of the $(x,y)$ plane is covered by initial conditions of orbits which escape entering the exterior region. Furthermore it is seen that the stability islands at the right side of the Moon is no longer present in this energy range. As the value of $z_0$ increases three important phenomena take place in the configuration space: (i) the area of the stability island located at the left side of the Moon decreases, (ii) the exterior realm basin, both with smooth and non-smooth boundaries increases, as well as, the region of fractal boundaries between the three types of basins grows, (iii) at the highest value of $z_0$ studied, that is $z_0 = 0.12$ (see Fig. \ref{hr3}d), both collision basins merge together. It should be pointed out that in this energy region we did not encountered the phenomenon of trapped chaos observed in the previous energy range. After integrating many sets of initial conditions for various values of the Jacobi constant and for several values of $z_0$ we conclude that in this energy range escape orbits to both Earth realm and exterior realm are always present regardless of how high is the value of $z_0$. In Fig. \ref{hr3t}(a-d) we illustrate the corresponding distribution of the escape and collision time of orbits on the configuration $(x,y)$ space.

\begin{figure*}
\centering
\resizebox{\hsize}{!}{\includegraphics{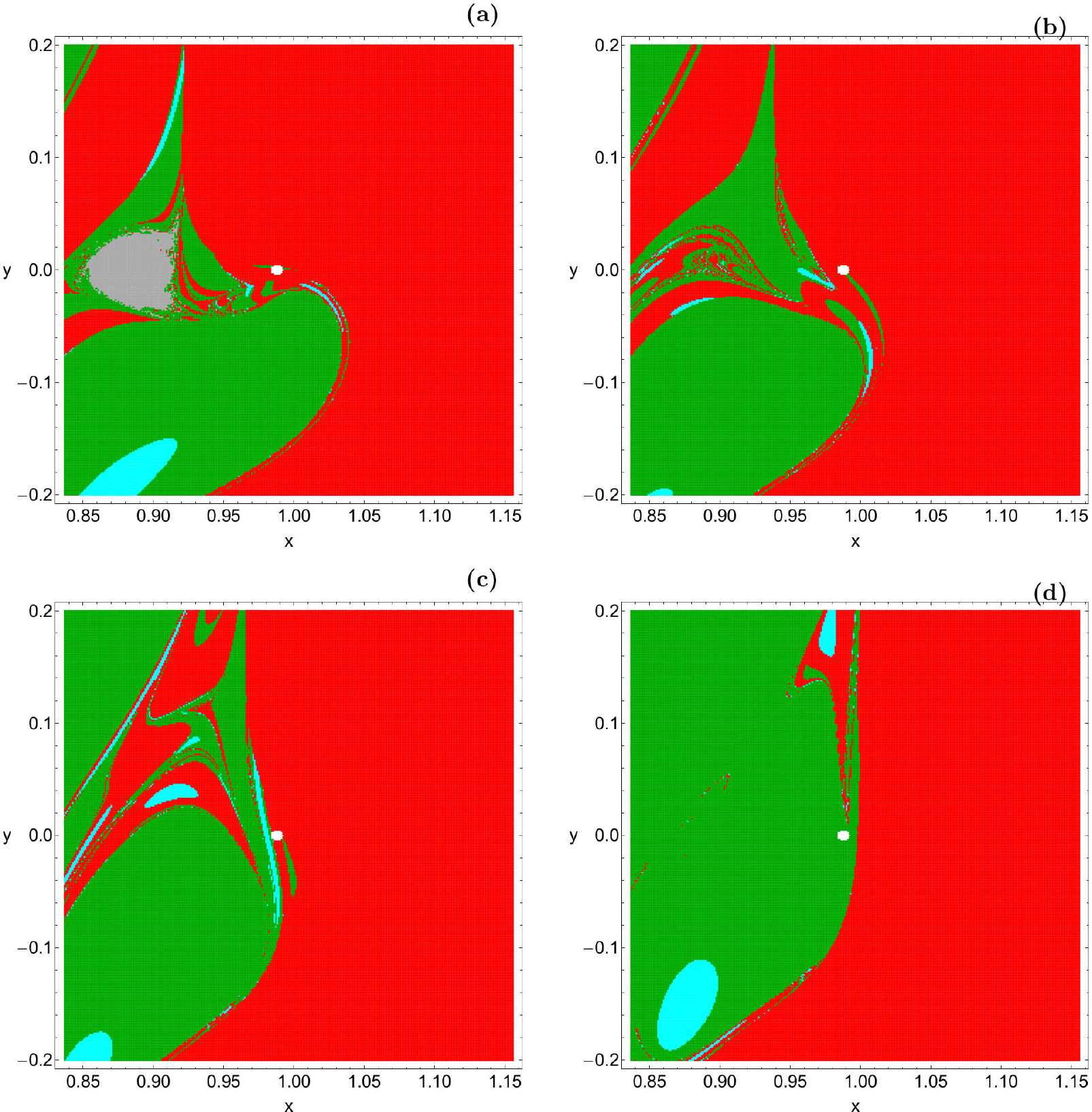}}
\caption{Basin diagrams for energy region III when $C = 3.012$. (a-upper left): $z_0 = 0.04$, (b-upper right): $z_0 = 0.09$, (c-lower left): $z_0 = 0.14$ and (d-lower right): $z_0 = 0.19$. The color code is as in Fig. \ref{hr3}.}
\label{hr4}
\end{figure*}

\begin{figure*}
\centering
\resizebox{\hsize}{!}{\includegraphics{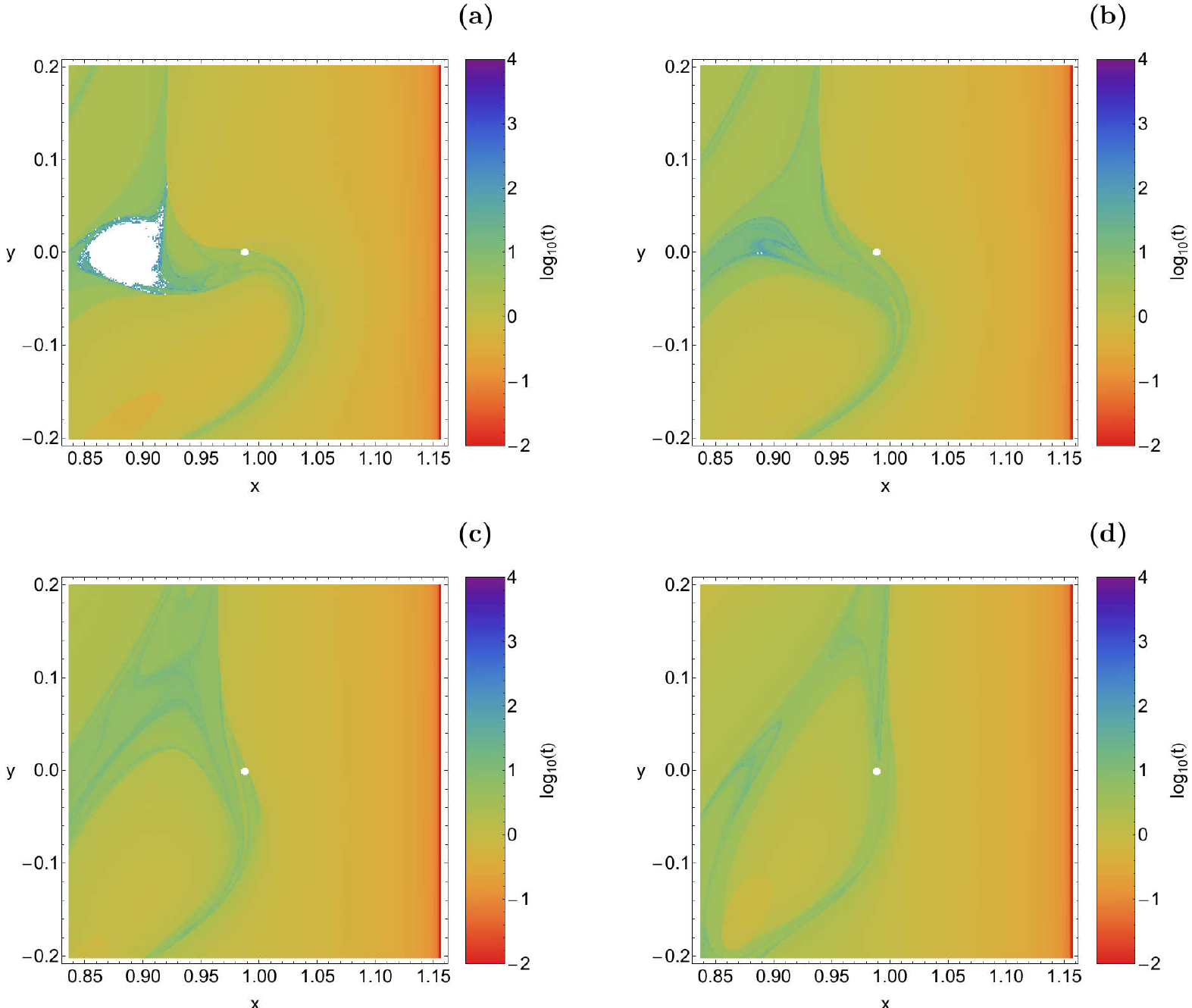}}
\caption{Distribution of the collision and escape time of the orbits on the configuration $(x,y)$ space for the values of $z_0$ of Fig. \ref{hr4}(a-d).}
\label{hr4t}
\end{figure*}

\begin{figure*}
\centering
\resizebox{\hsize}{!}{\includegraphics{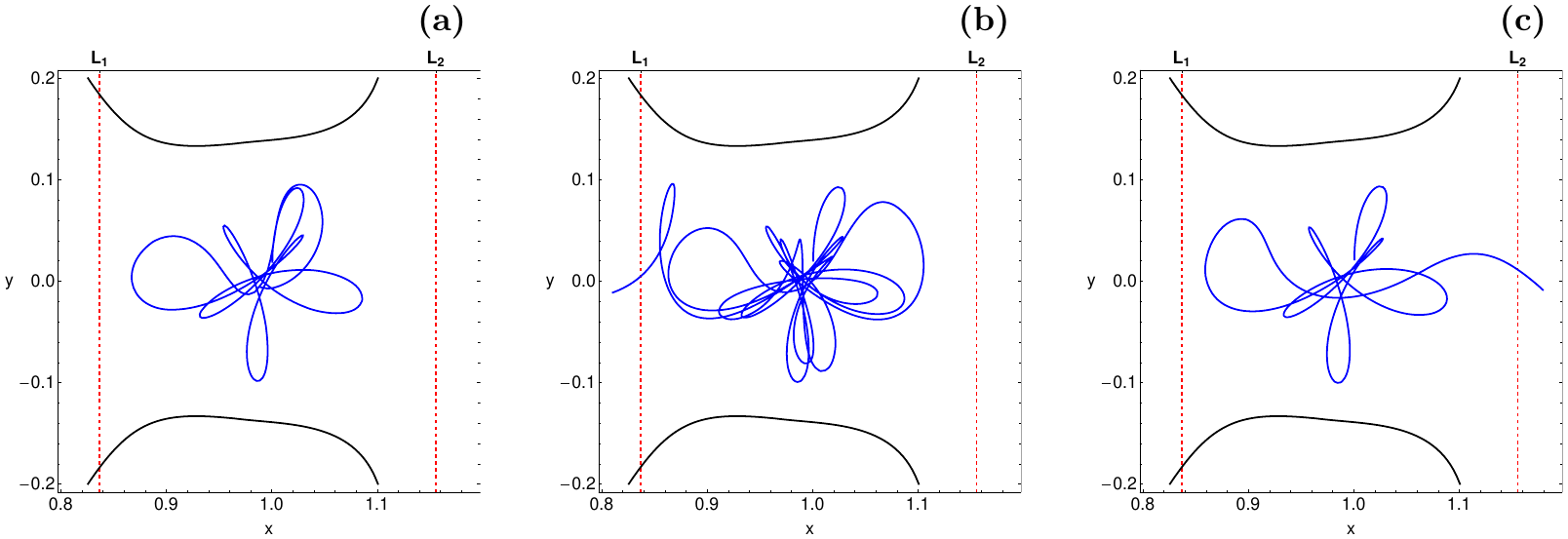}}
\caption{Orbits with different initial condition $y_0$ which lead to different type of motion. The outermost black solid line is the ZVC, while the position of the Lagrange points $L_1$ and $L_2$ is indicated by vertical, dashed, red lines. More details about the orbits are given in the text.}
\label{orbs}
\end{figure*}

\subsection{Energy region IV: $C_3 > C > C_4$}
\label{ss4}

When $C < C_3$ the neck around $L_3$ opens thus allowing orbits to escape to the exterior region also from the left side of the Earth. Since we decided to focus our study in the vicinity of the Moon the new escape channel in the ZVC is not visible this time. In Fig. \ref{hr4}(a-d) we observe for four values of $z_0$ the orbital structure of the configuration space when $C = 3.012$. In Fig. \ref{hr4}a where $z_0 = 0.04$ we see that the majority of the configuration $(x,y)$ plane is covered by initial conditions that lead to escape to the exterior region. The second most populated type of initial conditions corresponds to the Earth realm basins. The stability island at the left side of the Moon and also the collision basins are still present. When $z_0 = 0.09$ one may observe that the stability island has disappeared, while the collision basins are very limited and located mainly in the boundaries between the escape basins. For all larger and permissible values of $z_0$ the escape basins dominate the configuration space. In fact we may say that the divide the $(x,y)$ plane into two areas around the Moon. In particular the majority of the left side of the Moon is occupied by initial conditions which lead to escape to Earth realm, while almost all the right side of the Moon is covered by initial conditions of orbits which escape to the exterior region. For $z_0 = 0.19$ it is seen in Fig. \ref{hr4}d that a small collision basin emerges inside the Earth realm basin. The distribution of the escape and collision times of orbits on the configuration space is depicted in Fig. \ref{hr4t}(a-d). We see similar outcomes with that presented in the three previous subsections.

Both fractal and non-fractal (smooth) basin boundaries have been identified in the color-coded OTDs shown in Figs. \ref{hr1}, \ref{hr2}, \ref{hr3} and \ref{hr4}. The existence of fractal basin boundaries is a very common phenomenon observed in leaking Hamiltonian systems \citep[e.g.,][]{BGOB88,dML99,dMG02,STN02,ST03,TSPT04}. In the CRTBP system the leakages can be defined by both collisional and escaping initial conditions of orbits. The high complexity of the basin boundaries implies that in these regions is very difficult, or even impossible, to predict the type of motion of the test particle (e.g., a satellite, asteroid, planet etc). In Fig. \ref{orbs}(a-c) we present an illuminating example showing the highly unpredictability in the fractal regions of the plots. The initial conditions of the three orbits were taken form the OTD of Fig. \ref{hr3}c where $C = 3.104$. All three orbits have: $x_0 = 1.00075072$, $z_0 = 0.09$, $\dot{x_0} = \dot{z_0} = 0$, $\dot{y_0} > 0$ (obtained from the Jacobi integral of motion), while only $y_0$ is different. In particular when $y_0 = 0.0195$ we see in Fig. \ref{orbs}a that the corresponding orbit collides with the surface of the Moon after about 6.8 dtu. On the other hand, when $y_0 = 0.0196$ (panel b) and $y_0 = 0.0197$ (panel c) the test particle escapes to the Earth and to the exterior realm after about 14.5 dtu and 6.7 dtu, respectively. Thus it becomes evident that inside the fractal regions a slight change to the initial conditions of an orbit leads to completely different type of motion. Similar orbital behavior applies to all fractal regions of the OTDs.

\begin{figure*}
\centering
\resizebox{\hsize}{!}{\includegraphics{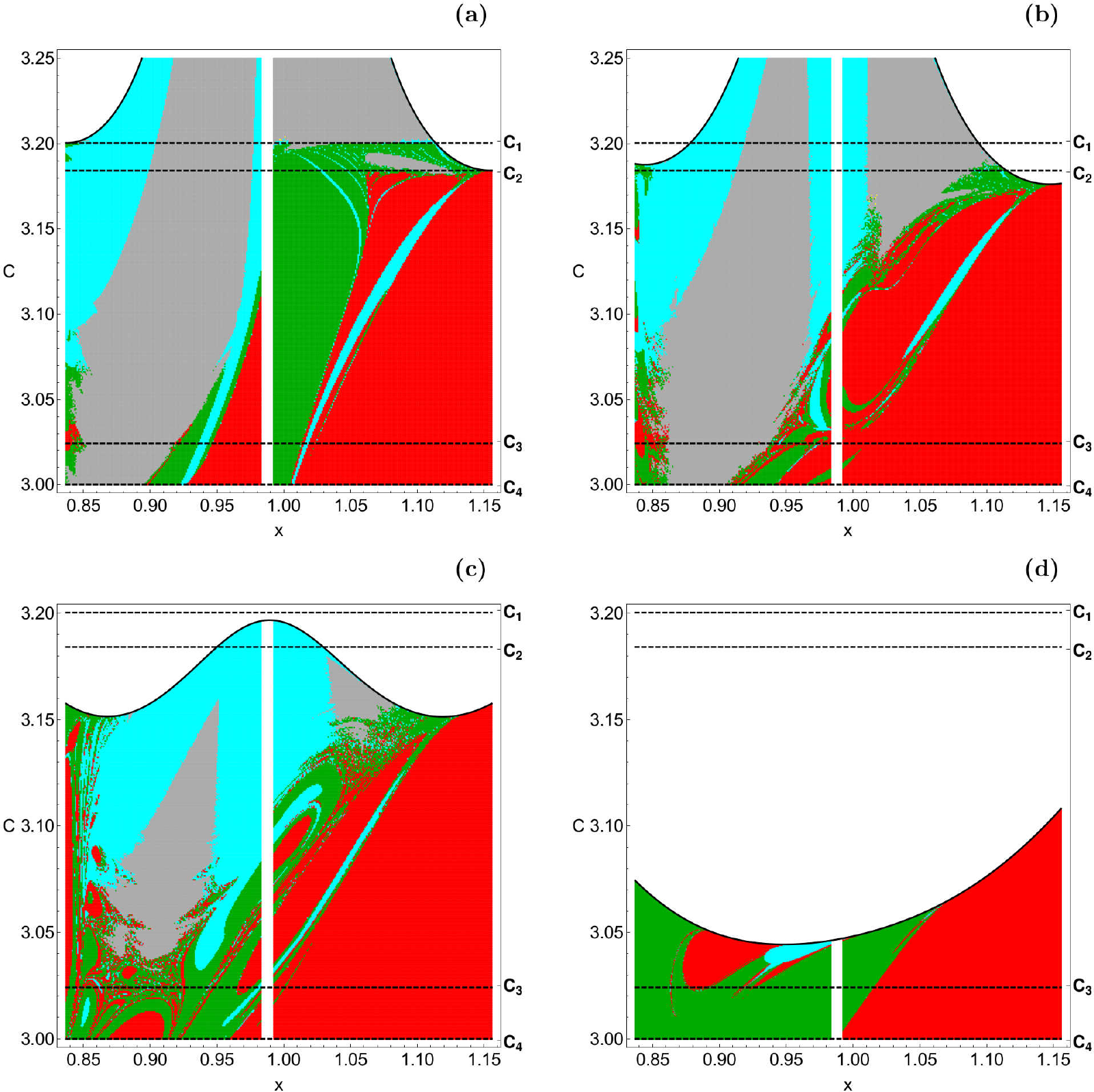}}
\caption{Orbital structure of the $(x,C)$ plane when (a-upper left): $z_0 = 0.001$; (b-upper right): $z_0 = 0.05$; (c-lower left): $z_0 = 0.1$; (d-lower right): $z_0 = 0.2$. The color code is the same as in Fig. \ref{hr3}.}
\label{xc}
\end{figure*}

\begin{figure*}
\centering
\resizebox{\hsize}{!}{\includegraphics{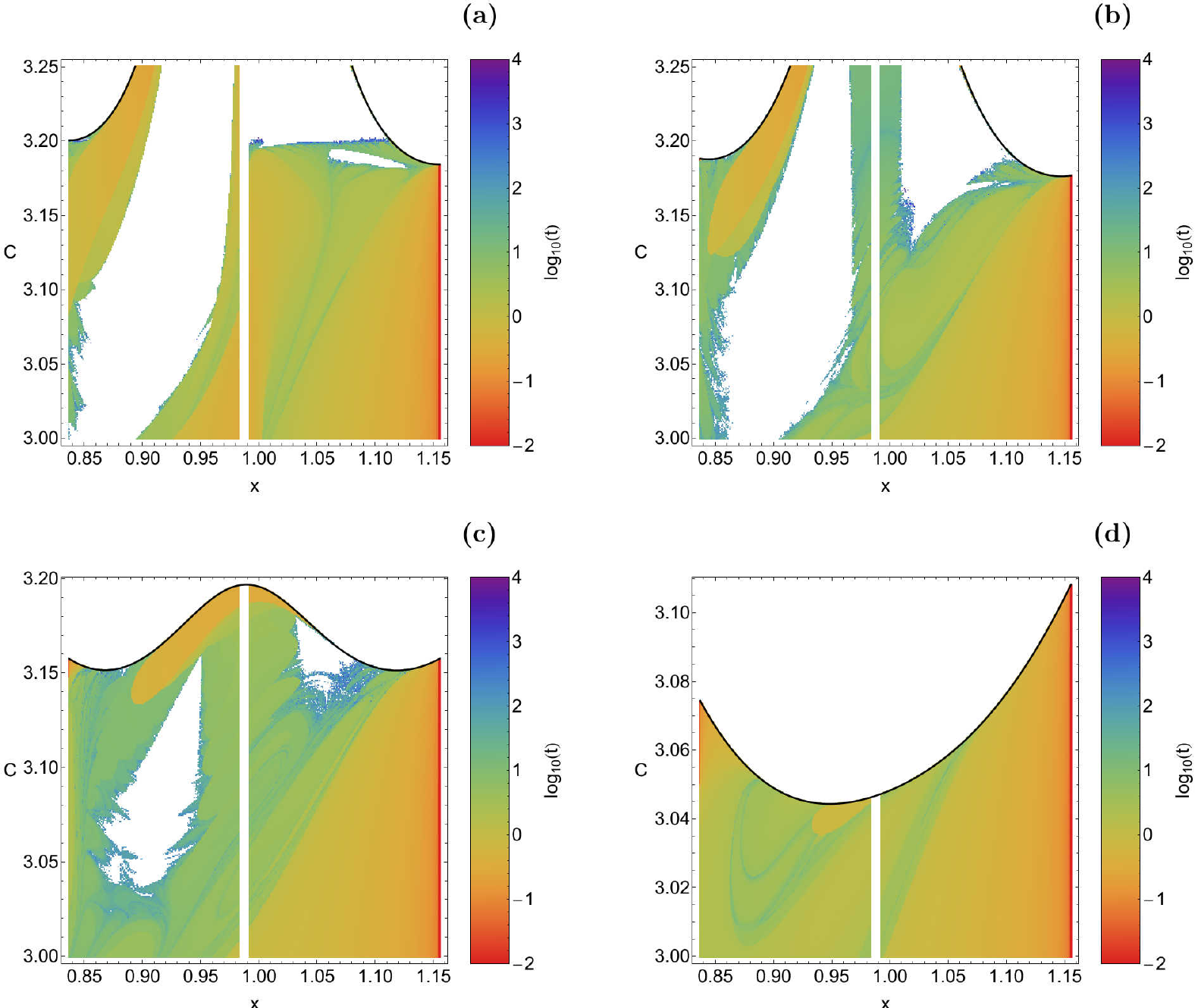}}
\caption{The distribution of the corresponding escape and collision times of the orbits of the $(x,C)$ planes shown in Fig. \ref{xc}.}
\label{xct}
\end{figure*}

\begin{figure*}
\resizebox{\hsize}{!}{\includegraphics{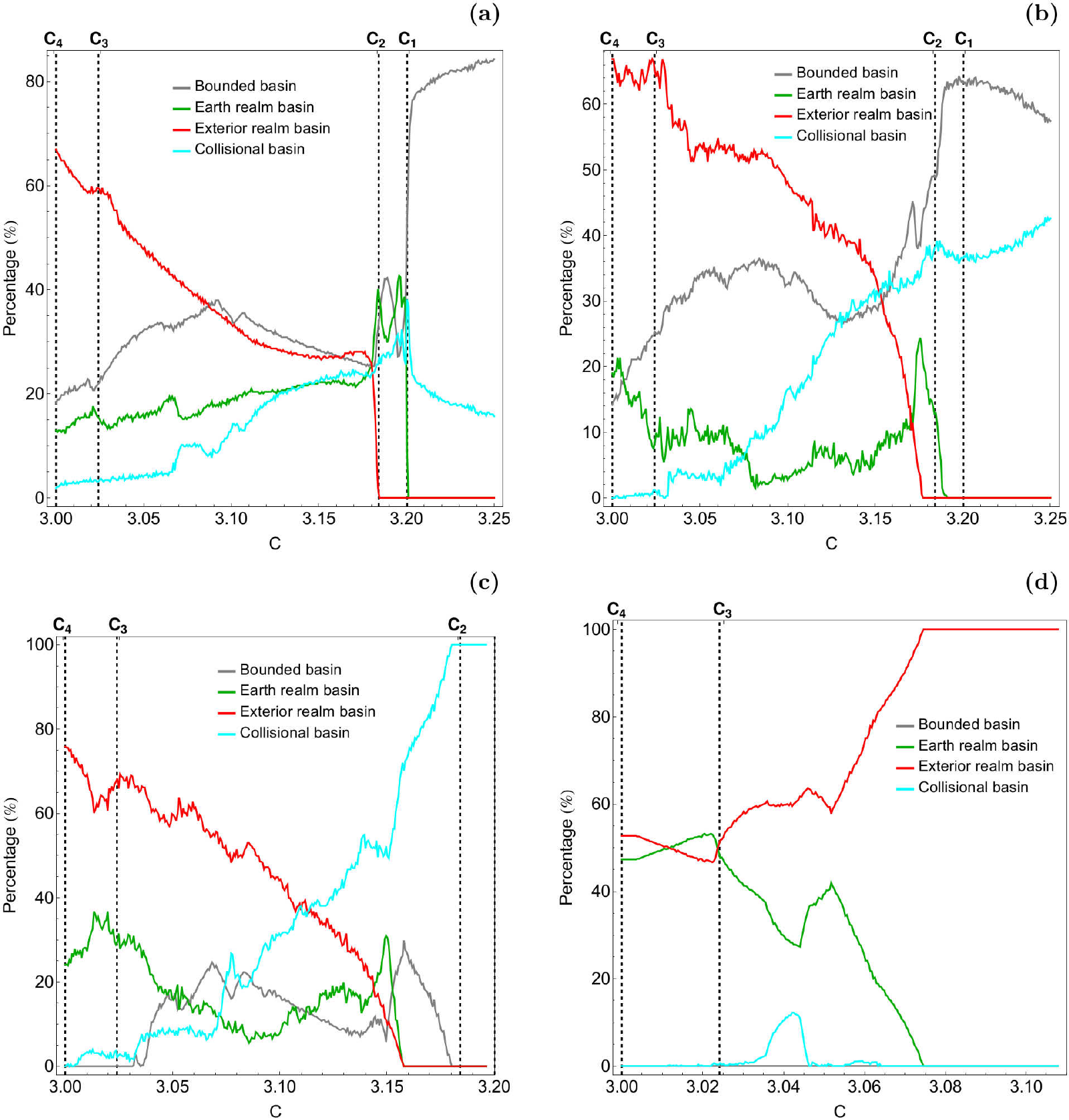}}
\caption{Evolution of the percentages of the initial conditions of each considered basin as a function of the Jacobi constant when (a-upper left): $z_0 = 0.001$; (b-upper right): $z_0 = 0.05$; (c-lower left): $z_0 = 0.1$; (d-lower right): $z_0 = 0.2$. The vertical dashed black lines indicate the four critical values of $C$.}
\label{p1}
\end{figure*}

\begin{figure*}
\centering
\resizebox{\hsize}{!}{\includegraphics{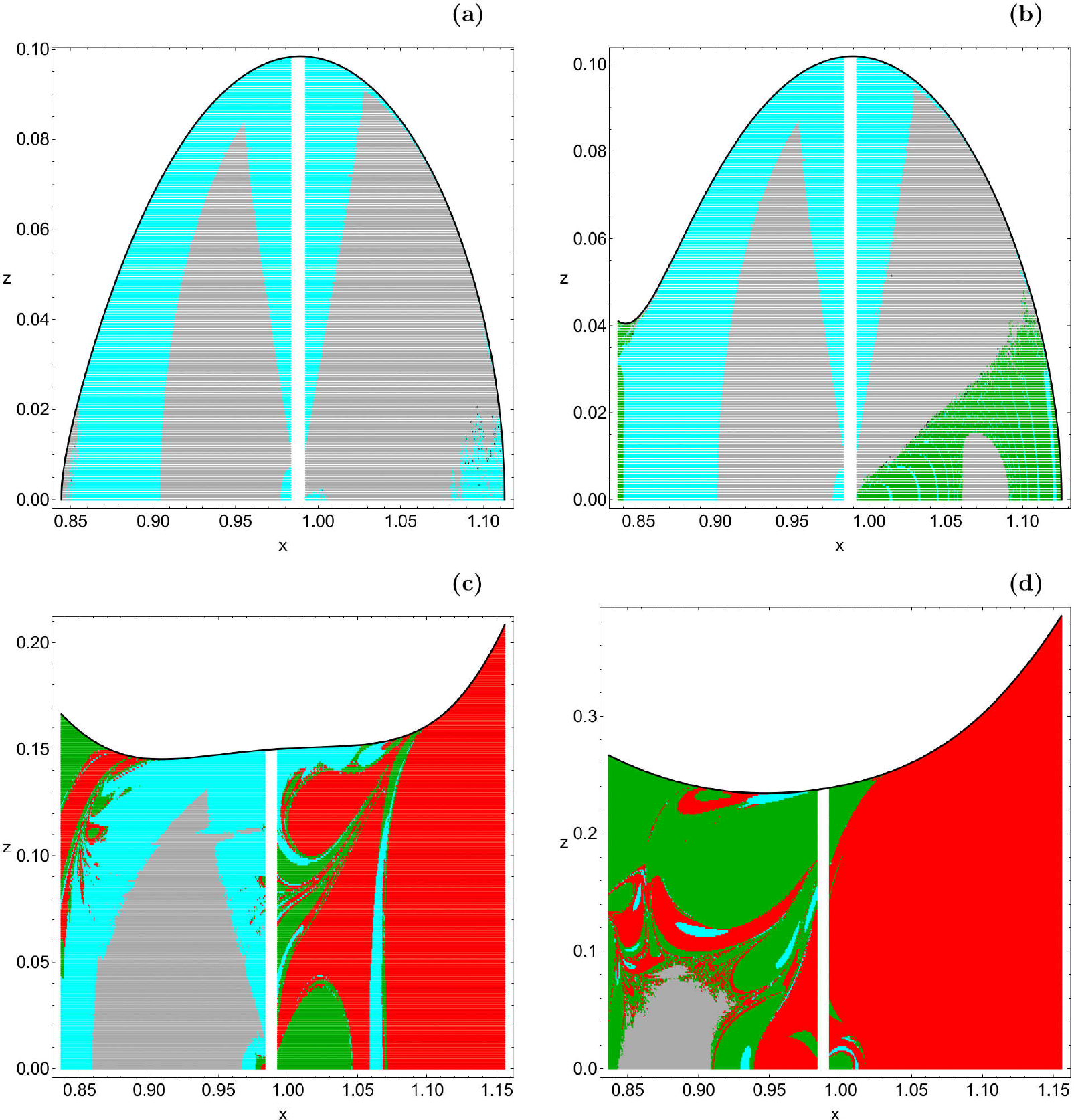}}
\caption{Orbital structure of the $(x,z)$ plane when (a-upper left): $C = 3.201$; (b-upper right): $C = 3.192$; (c-lower left): $C = 3.104$; (d-lower right): $C = 3.012$. The color code is the same as in Fig. \ref{hr3}.}
\label{xz}
\end{figure*}

\begin{figure*}
\centering
\resizebox{\hsize}{!}{\includegraphics{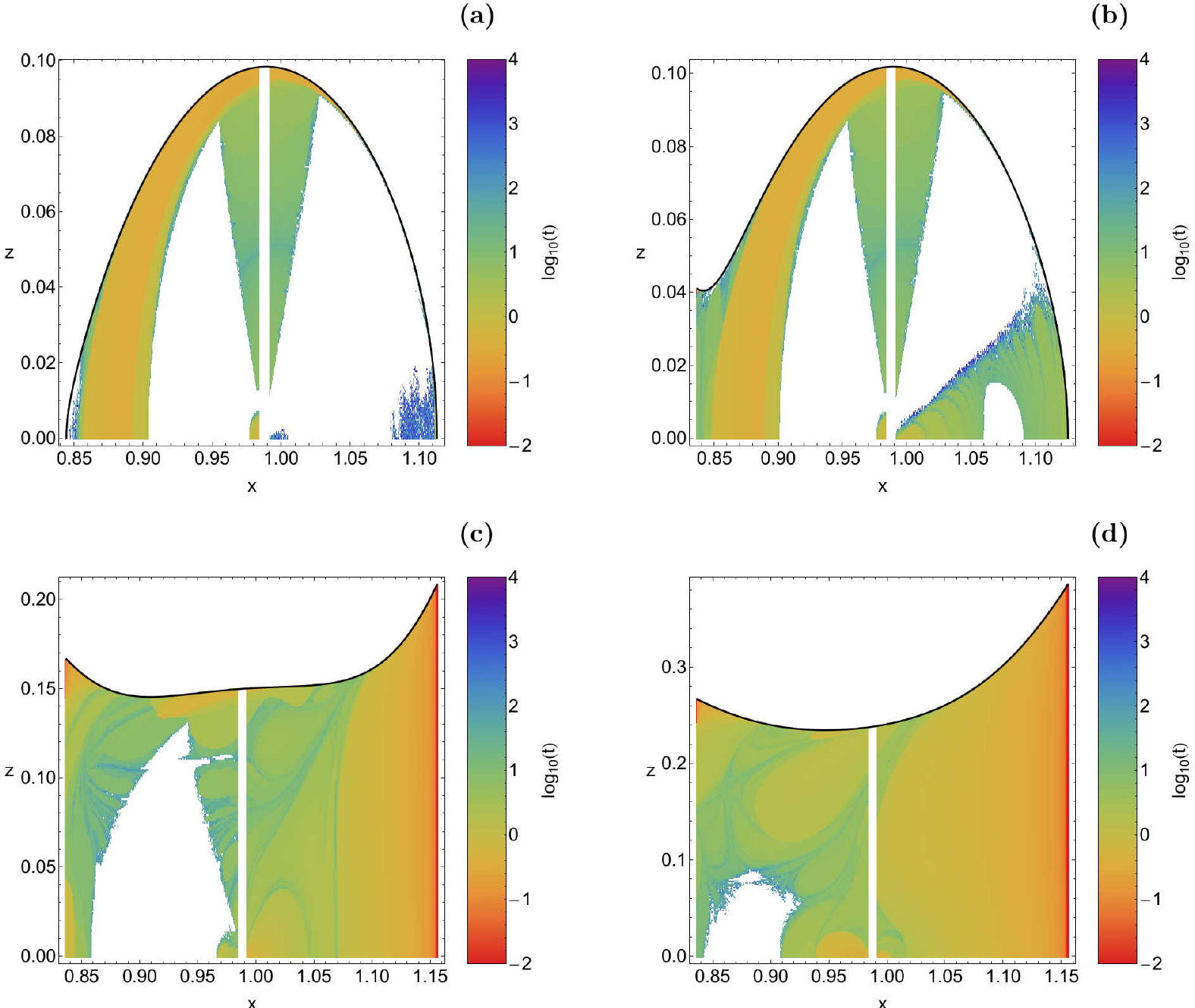}}
\caption{The distribution of the corresponding escape and collision times of the orbits of the $(x,z)$ planes shown in Fig. \ref{xz}.}
\label{xzt}
\end{figure*}

\begin{figure*}
\resizebox{\hsize}{!}{\includegraphics{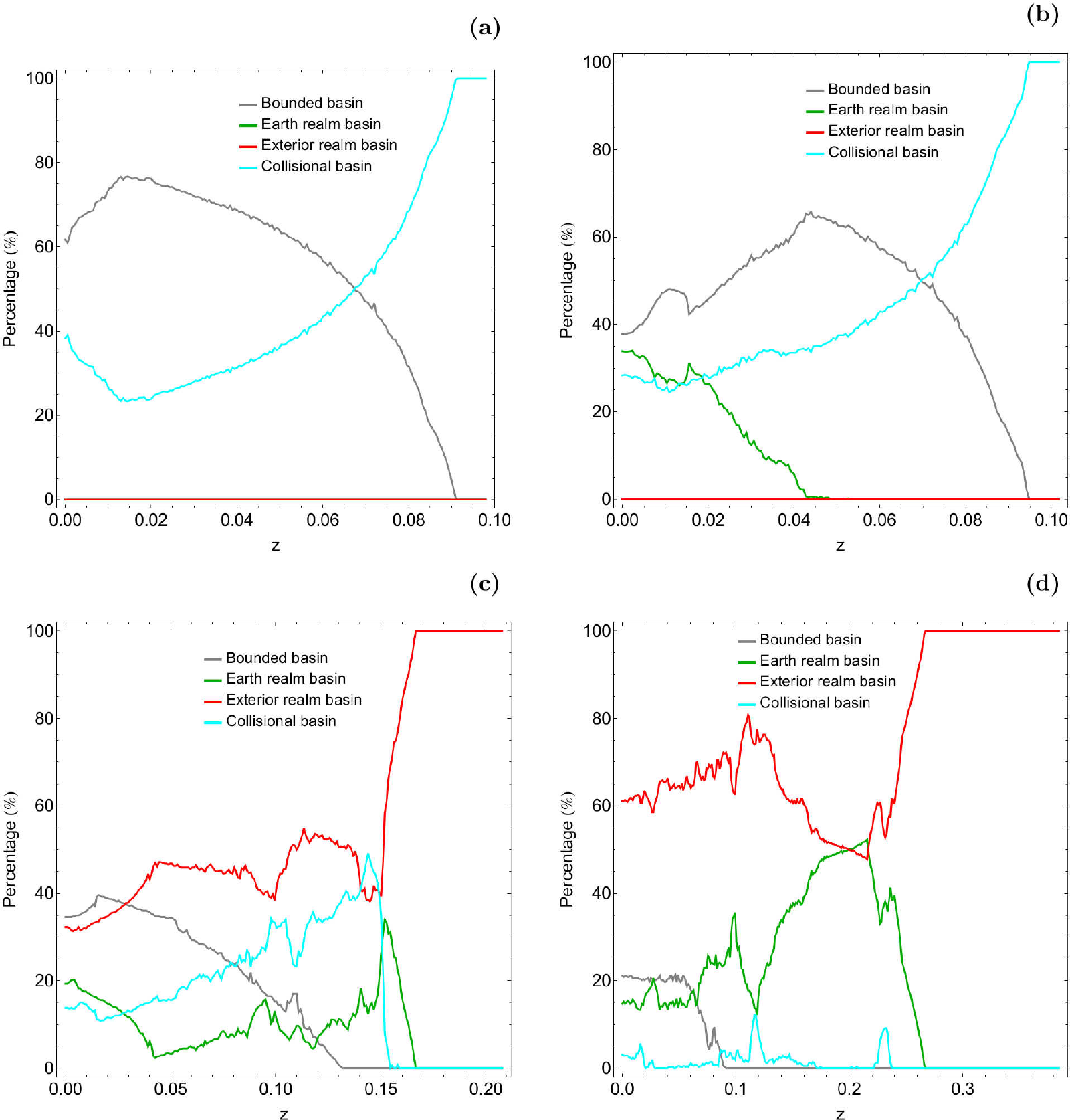}}
\caption{Evolution of the percentages of the initial conditions of each considered basin as a function of the initial value of the $z$ coordinate when (a-upper left): $C = 3.201$; (b-upper right): $C = 3.192$; (c-lower left): $C = 3.104$; (d-lower right): $C = 3.012$.}
\label{p2}
\end{figure*}

\subsection{An overview analysis}
\label{over}

Using the color-coded OTDs in the configuration $(x,y)$ space we can obtain sufficient information on the phase space mixing however, for only a fixed value of the Jacobi constant and also for orbits that traverse the surface of section either directly (progradely) or retrogradely. H\'{e}non \citep{H69}, in order to overcome these limitations introduced a new type of plane which can provide information regarding the orbital structure of the system using the section $y = \dot{x} = \dot{z} = 0$, $\dot{y} > 0$ \citep[see also][]{BBS08}. In other words, all the initial conditions of the 3D orbits are launched from the $x$-axis with $x = x_0$, parallel to the $y$-axis $(y = 0)$ with a specific value of $z_0$. Consequently, in contrast to the previously discussed OTDs, only orbits with pericenters on the $x$-axis are included and therefore, the value of the Jacobi constant $C$ can now be used as an ordinate. In this way, we can monitor how the variation on $C$ influences the overall orbital structure of the Earth-Moon system using a continuous spectrum of Jacobi constant values rather than few discrete levels. In Fig. \ref{xc}(a-d) we present, for four values of $z_0$, the orbital structure of the $(x,C)$ plane when $C \in [C_4,3.25]$. The four critical values of $C$ are indicated by the horizontal dashed black lines. The outermost black solid line in Fig. \ref{xc}(a-d) is the limiting curve which distinguishes between regions of allowed and forbidden motion and is defined as
\begin{equation}
f_1(x,C;z_0) = 2\Omega(x,y = 0,z = z_0) = C.
\label{zvc1}
\end{equation}

We begin our exploration with $z_0 = 0.001$, which is a $z_0$ value very close to the $(x,y)$ plane. This suggests that the results should be very similar to those obtained in Paper I for the two-dimensional system $(z = 0)$. Indeed in Fig. \ref{xc}a we see that the orbital structure of the $(x,C)$ plane is quite similar to that presented in Fig. 19 of Paper I. All types of motion are present forming several basins. In most cases the boundaries between the basins are very smooth however there are also areas with fractal structures. In particular, it is seen that the fractality of the basin boundaries migrates from the right side of the Moon for high values of $C$ (low energy levels) to the left of the same primary for low values of $C$ (high energy levels). We then proceed to $z_0 = 0.05$ and our results are shown in Fig. \ref{xc}b. We observe that the increase of the initial value of the $z$ coordinate leads to several changes to the $(x,C)$ plane. The most important ones are the following: (i) the stability island on the right side of the Moon extends below $C_2$, (ii) the basins corresponding to the Earth realm on the right side of the Moon heavily reduce, (iii) the fractality on several areas of the $(x,C)$ plane increases. For $z_0 = 0.1$ one may observe in Fig. \ref{xc}c a very interesting orbital structure. The size of both stability islands is reduced, while on the left side of the Moon there is a highly fractal mixture of escape and collision basins. The pattern on the $(x,C)$ plane changes drastically in Fig. \ref{xc}d where $z_0 = 0.2$. Here there is no indication of regular bounded motion whatsoever, while the fractal regions are almost negligible. Note that the permissible area on the $(x,C)$ plane reduces with increasing $z_0$. In Fig. \ref{xct}(a-d) we illustrate how the corresponding escape and collision times of orbits are distributed on the $(x,C)$ plane.

Useful conclusions can be obtained by monitoring the evolution of the percentages of all types of orbits as a function of the Jacobi constant $C$. Fig. \ref{p1}(a-d) shows the diagrams corresponding to the $(x,C)$ planes of Fig. \ref{xc}(a-d), respectively. The vertical dashed black lines indicate the four critical values of the Jacobi constant. We observe that in each case (value of $z_0$) the evolution of the respective percentages is very different. For low values of $z_0$ $(z_0 \leq 0.05)$ bounded basins of regular motion dominate when $C > C_1$, while on the other hand for high values of $z_0$ $(z_0 > 0.05)$ there is no indication of bounded motion in the same energy region since the corresponding percentage is zero. In particular, for $z_0 = 0.2$ bounded motion is completely absent in the entire energy range studied. In the energy range between $C_2$ and $C_3$ the rate of bounded orbits exhibit fluctuations, while for $C < C_3$ it reduces or even vanishes for $z_0 \geq 0.1$. For high values of the Jacobi constant $(C > C_2)$, or in other words for low energy levels, collision basins occupy almost all the available space for $z_0 = 0.1$, while for $C < C_2$ the rate of collisional orbits starts to decline and for $C < C_4$ it vanishes. For lower values of $z_0$ the evolution of the same type of orbits is very similar in the energy range $C < C_2$. On the contrary for $z_0 = 0.2$ the percentage of collisional orbits is almost zero in the majority of the energy ranges except around $C = 3.04$ where a peak at about 10\% is observed. As soon as the transport channel to the Earth realm opens, for $C < C_1$, the percentage of escaping orbits to the Earth realm starts to grow in all cases displaying fluctuations in the energy range between $C_1$ and $C_4$. It is interesting to notice that the percentage corresponding to the Earth realm does not appear in all cases at the same value of $C$. Being more precise as the value of $z_0$ increases escaping orbits to Earth realm appear for the first time in lower values of $C$. In the same vein, the rate of escaping orbits to the exterior realm appear for $C < C_2$ and once more the energy level of their first indication is related to the value of $z_0$. For $z_0 \leq 0.1$ the percentage of escaping orbits to the exterior realm rapidly grows occupying more than 50\% of the available space for $C < C_4$. However for $z_0 = 0.2$ the exterior realm basins cover all the available space for $C > 3.075$, while for lower values of the Jacobi constant the same rate drops and for $C < C_4$ it seems to saturate at about 55\%. Here we would like to clarify that we did not included the evolution of trapped chaotic orbits in the diagrams because the corresponding percentage is extremely low (lower than 0.5\%) throughout the energy range studied.

It would be very illuminating if we had a more complete view of how the initial value of $z_0$ influences the nature of orbits of the Earth-Moon system. In order to obtain this we follow a similar numerical approach to that explained before for the Jacobi constant thus examining now a continuous spectrum of $z_0$ values. In particular, we use again the section $y = \dot{x} = \dot{z} = 0$, $\dot{y} > 0$, launching orbits once more from the $x$-axis with $x = x_0$, parallel to the $y$-axis, with $z = z_0$. This allow us to construct again a two-dimensional plane in which the $x$ coordinate of orbits is the abscissa, while the $z$ coordinate is the ordinate. In Fig. \ref{xz}(a-d) we present the orbital structure of the $(x,z)$-plane when $z \in [0,z_{max}]$ (the 2D case where $z_0 = 0$ is also included), for four values of the Jacobi constant, while in Fig. \ref{xzt}(a-d) the distribution of the corresponding escape and collision times of orbits is depicted. The maximum allowed value of the $z$ coordinate, $z_{max}$, is related with the particular value of the energy level. The outermost black solid line is the limiting curve which this time is given by
\begin{equation}
f_2(x,z;C) = 2\Omega(x,y = 0,z) = C.
\label{zvc2}
\end{equation}

For $C = 3.201$, that is an energy level which corresponds to the first Hill's regions configurations, we see in Fig. \ref{xz}a that bounded basins and collision basins share the $(x,z)$ plane. For relative high values of $z_0$ however only collisional motion is possible. In Fig. \ref{xz}b where $C = 3.192$ it is seen that at the right side of the Moon an Earth realm basin is present. This plot reveals the interesting phenomenon discussed earlier in subsection \ref{ss2} where we observed that for relative high values of $z_0$ orbits do not enter the Earth realm, even though the corresponding transport channel is open. Indeed in Fig. \ref{xz}b it is evident that the Earth realm basin ends at about $z_0 = 0.04$ which coincides with the height of the transport channel. When $C = 3.104$ the stability island on the right side of the Moon disappears, while as we can see in Fig. \ref{xz}c the same area is covered mainly by escape basins. Furthermore collision basins are mostly observed at the opposite side of the $(x,z)$ plane. The extent of the stability island shrinks when $C = 3.012$ as it is seen in Fig. \ref{xz}d. The right side of the Moon is almost entirely covered by escaping orbits to the exterior region, while the left side of the same primary contains a mixture of escaping and collisional orbits. We see that at this energy case (Hill's regions configurations) escape basins dominate, while the collision basins are very confined with respect to the previous energy case.

We close our numerical investigation by presenting the evolution of the percentages of all types of orbits as a function of the initial value of the $z$ coordinate $(z_0)$. Fig. \ref{p2}(a-d) contains the diagrams corresponding to the $(x,z)$ planes of Fig. \ref{xz}(a-d), respectively. In Fig. \ref{p2}a we see that for about $z_0 < 0.067$ bounded regular orbits is the most populated type of orbits, while for about $z_0 > 0.067$ collisional orbits dominate. The same applies also for $C = 3.192$ as we can see in Fig. \ref{p2}b. Here we observe that the percentage of escaping orbits to the Earth realm starts at about 35\% when $z_0 = 0$, however the rate reduces with increasing $z_0$. In fact for $z_0 > 0.04$ there is no indication of escaping orbits whatsoever. When $C = 3.104$ it is seen in Fig. \ref{p2}c that the percentage of escaping orbits to the exterior realm prevails almost in the entire energy range. In the previous two energy levels (see panels a and b of Fig. \ref{p2}) we observed that for relative high values of $z_0$ collisional motion dominates. In this energy level however this is not true. In particular for about $z_0 > 0.15$ both escaping orbits to the Earth realm and collisional orbits disappear, while at the same time the rate of escaping orbits to the exterior realm grows rapidly and for about $z_0 > 0.165$ this type of orbits occupy the entire available space. When $C = 3.012$, which corresponds to the fourth Hill's regions configuration, bounded regular orbits vanish more quickly for $z_0 > 0.1$, while the rate of collisional orbits is very low (less than 5\%) throughout except for some peaks. Once more for $z_0 > 0.22$ the rate of escaping orbits to the Earth realm is reduced, while that of escaping orbits to the exterior realm increases. Thus taking into account all the above-mentioned analysis we may conclude that in the first and second Hill's regions configurations collisional motion dominates for high values of $z_0$, while on the other hand in the third and fourth Hill's regions configurations escaping orbits to the exterior realm is the most populated orbital family for relative high initial values of the $z$ coordinate. Once more, as in Fig. \ref{p1} we did not included the evolution of the percentage of trapped chaotic orbits because it is almost negligible (lower than 0.1\%) throughout the range of the values of the $z$ coordinate.

\begin{figure*}
\resizebox{\hsize}{!}{\includegraphics{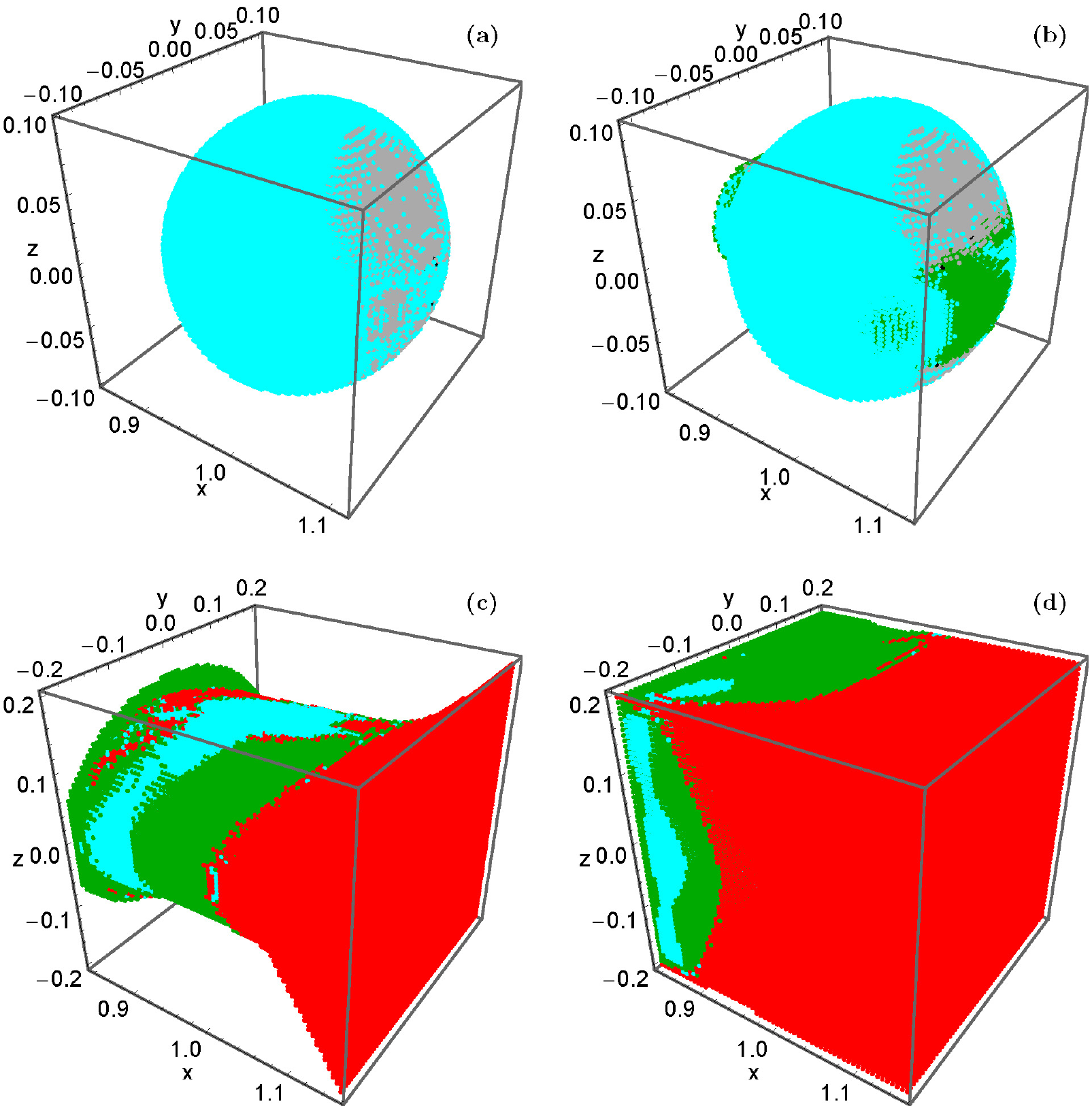}}
\caption{Orbital structure of three dimensional distributions of initial conditions of orbits in the configuration $(x,y,z)$ space when (a-upper left): $C = 3.201$; (b-upper right): $C = 3.192$; (c-lower left): $C = 3.104$; (d-lower right): $C = 3.012$.}
\label{3dgr}
\end{figure*}

\subsection{Three dimensional distributions of initial conditions of orbits}
\label{3d}

In all previous subsections we investigated the orbital dynamics of orbits using two dimensional grids of initial conditions in several types of planes. Taking into account that the scope of this work is to explore the three dimensional version of the Earth-Moon system, we decided to expand our search using three dimensional grids of initial conditions. Being more precise, for a particular value of the energy we define inside the corresponding zero velocity surface (see Fig. \ref{hrcs}) dense uniform grids of $100 \times 100 \times 100$ initial conditions $(x_0,y_0,z_0)$ with $\dot{x_0} = \dot{z_0} = 0$, while the initial value of $\dot{y} > 0$ is always obtained from the Jacobi integral of motion (\ref{ham}). The scattering region is defined as: $x_{L_1} \leq x \leq x_{L_2}$ and $-0.2 \leq y,z \leq 0.2$.

In Fig. \ref{3dgr}(a-d) we present the orbital structure of the three dimensional distributions of initial conditions of orbits in the configuration $(x,y,z)$ space for four characteristic energy levels corresponding to the first four Hill's regions configurations. The color code is the same as in Fig. \ref{hr3}. It is seen that in this case the grid of initial conditions is in fact a three dimensional solid and therefore only its outer surface is visible. A tomographic-style approach can be used in order to penetrate and examine the interior region of the solid \citep[e.g.,][]{Z14a}. This task however, exceeds the scope of the present paper. Nevertheless, in previous subsections \ref{ss1} - \ref{ss4} we classified initial conditions of orbits in several levels of $z_0$ (see Figs. \ref{hr1}, \ref{hr2}, \ref{hr3}, \ref{hr4}) which can be considered as horizontal tomographic slices of the corresponding three dimensional solids shown in Fig. \ref{3dgr}. Thus we may say that we obtained a spherical view of the orbital dynamics of the three dimensional Earth-Moon system. Here we would like to point out that to our knowledge this is the first time that a three dimensional distribution of initial conditions is used for classifying orbits in the CRTBP.

\begin{figure}
\resizebox{\hsize}{!}{\includegraphics{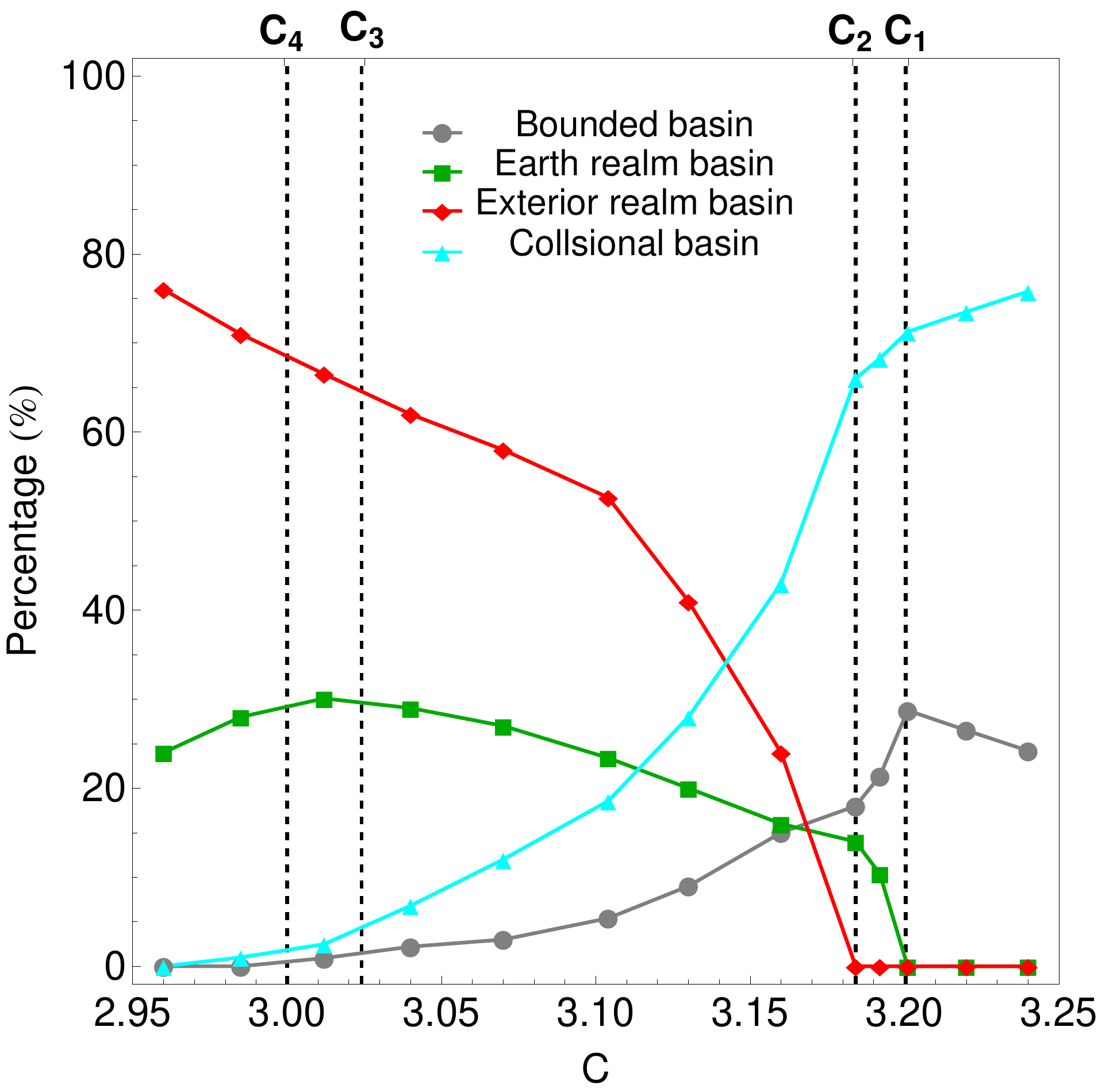}}
\caption{Evolution of the percentages of all types of orbits with initial conditions inside the three dimensional grids, as a function of the Jacobi constant $C$.}
\label{percs}
\end{figure}

Using the above-mentioned technique regarding the three dimensional distribution of initial conditions of orbits we can monitor how the percentages of the different types of orbits in the 3D configuration space evolve as a function of the Jacobi constant $C$. For obtaining a more complete view we integrated more 3D grids (not shown here) for additional energy levels. Our results are presented in Fig. \ref{percs}. We see that for high values of the Jacobi constant (or equivalently for low energy levels), that is for $C > C_1$, collisional motion dominates occupying more than 70\% of the configuration space. However the rate of collisional orbits is reduced with increasing energy and for $C < C_4$ it vanishes. The evolution of the percentage of bounded regular orbits is very similar. Bounded regular motion corresponds to about 25\% of the $(x,y,z)$ space for $C > C_1$, while for lower values of the Jacobi constant its rate is reduced and for $C < C_4$ it completely disappears. The percentages of escaping orbits (to both realms) exhibit a different pattern. In particular, one may observe that the rate of escaping orbits to the Earth realm appears as soon as $C < C_1$, and it grows for higher values of the total orbital energy until $C = C_4$, where its starts to decline. Our calculations reveal that the percentage of the Earth realm basins does not exceed 30\% throughout the energy range. For $C < C_2$ the neck near $L_2$ opens thus allowing orbits to escape to the exterior region. At the same time the corresponding percentage grows rapidly and for $C < 3.14$ escaping orbits to the exterior realm is the most populated type of orbits. The computations suggest that for $C < 2.95$ the exterior realm basins occupy more than 80\% of the configuration space.

\section{Discussion and conclusions}
\label{disc}

We used the three dimensional version of the circular restricted three-body problem (CRTBP) in order to numerically investigate the orbital dynamics of a small body (spacecraft, comet or asteroid) under the influence of the potential of the Earth-Moon system. In particular, we expanded into three dimensions the numerical analysis of Paper I regarding the planar Earth-Moon system. All the initial conditions of orbits were initiated in the vicinity of the Moon which was our scattering region. We managed to determine the corresponding basins leading to five types of orbits: (i) bounded regular orbits around the Moon, (ii) trapped chaotic orbits around the Moon; (iii) escaping orbits to Earth realm through $L_1$, (iv) escaping orbits to the exterior realm through $L_2$, and (v) leaking orbits due to collisions with the surface of the Moon. The orbital structure of the basins was monitored by varying both the value of the Jacobi constant and the initial value of the $z$ coordinate. As far as we know, this is the first time that the three dimensional orbital content in the vicinity of the Moon, or in any other planetary system, is explored through orbit classification in such a detailed and systematic way and this is exactly the novelty and the contribution of the current work.

In our exploration we adopted the same numerical methods used in Paper I. For the numerical integration of the sets of initial conditions of orbits in each type of plane, we needed between about 0.5 hours and 16 hours of CPU time on a Quad-Core i7 2.4 GHz PC, depending of course on the escape and collision rates of orbits in each case. For each initial condition the maximum time of the numerical integration was set to be equal to $10^4$ time units however, when a test particle escaped or collided with the Moon the numerical integration was effectively ended and proceeded to the next available initial condition.

We provide quantitative information regarding the escape and collision dynamics in the CRTBP. The main outcomes of our numerical research can be summarized as follows:
\begin{enumerate}
  \item In the first Hill's regions configurations $(C > C_1)$ only bounded and collisional types of motion are possible. It was found that as we proceed to higher values of $z_0$ the size of the stability islands is reduced and collision basins dominate the configuration space.
  \item Escaping orbits to the Earth realm are added in the second Hill's regions configurations $(C_1 > C > C_2)$. In this energy range we observed a very interesting phenomenon related with the presence of escaping orbits. In particular, escaping orbits are present only for values of $z_0$ up to the height of the transport channel. For higher values of $z_0$ escaping orbits completely disappear, even though the transport channel in the three dimensional zero velocity surface is still open.
  \item In the third Hill's regions configurations $(C_2 > C > C_3)$ all possible types of orbits are present. Our numerical analysis suggests that the size of the bounded basins corresponding to regular orbits is reduced with increasing $z_0$. At the same time the configuration space is covered by a mixture of several types of basins with both smooth and fractal basin boundaries.
  \item Basins of bounded ordered orbits are not always present. Indeed in the fourth Hill's regions configurations the stability island on the left side of the Moon disappears for high values of $z_0$. Moreover, the escape basins share the configuration space, while the collision basins are heavily confined. We also observed that the fractality is considerably reduced with increasing $z_0$.
  \item Our investigation of the $(x,C)$ planes revealed that for low energy levels bounded motion dominated for $z_0 \leq 0.05$, collisional motion for intermediate values of $z_0$ (i.e., $z_0 = 0.1$), while for high values of $z_0$ orbits which escape to the exterior region is the most populated family. On the other hand, for high energy levels (low values of the Jacobi constant) the exterior realm basin is prevailing regardless the initial value of the $z$ coordinate.
  \item The analysis of the $(x,z)$ planes suggests that collisional motion dominates in the first and second Hill's regions configurations, for high values of $z_0$, while on the other hand escaping orbits to the exterior realm is the most populated orbital family in the third and fourth Hill's regions configurations for relative high initial values of the $z$ coordinate. In particular, when the escaping orbits to the exterior region dominate all other types of orbits completely disappear.
  \item We proved that that the escape and collision times of orbits are directly linked to the basins of escape and collision. In particular, inside the basins of escape/collision as well as relatively away from the fractal domains, the shortest escape/collision rates of the orbits had been measured. On the other hand, the longest escape/collision periods correspond to initial conditions of orbits either near the fractal boundaries between the escape/collision basins or in the vicinity of the stability islands.
\end{enumerate}

Judging by the detailed and novel outcomes of our systematic numerical investigation we may say that our task has been successfully completed. We hope that the present numerical analysis and the corresponding results to be useful in the field of escape dynamics in the CRTBP. It is in our future plans to use the outcomes of this work in order to investigate the Sun-Jupiter system and determine the final states of the Jupiter's moons, co-orbital objects and asteroids.

\section*{Acknowledgments}

I would like to express my warmest thanks to the anonymous referee for the careful reading of the manuscript and for all the apt suggestions and comments which allowed us to improve both the quality and the clarity of the paper.

\end{document}